\documentclass[a4paper, twocolumn, 10pt, superscriptaddress, pra, aps]{revtex4-1}
\usepackage[T1]{fontenc}
 \pdfoutput=1
\usepackage[utf8]{inputenc}
\usepackage{lmodern}

\usepackage{subfigure} 
\usepackage{graphicx} 
\usepackage{amsmath}
\usepackage{amsfonts} 
\usepackage{amssymb}  
\usepackage{mathtools}
\usepackage{xcolor} 
\usepackage{multirow}
\usepackage{comment}

\usepackage{color}   
\usepackage{hyperref}
\hypersetup{
	colorlinks=true, 
	linktoc=all,     
	linkcolor=blue,  
}

\newenvironment{lcase}
{\left\lbrace \begin{aligned}}
	{\end{aligned} \right.}

		{\end{pmatrix}\end{medsize}}%

\DeclarePairedDelimiter\abs{\lvert}{\rvert}%
\DeclarePairedDelimiter\ket{\lvert}{\rangle}%
\DeclarePairedDelimiter\bra{\langle}{\rvert}%
\DeclarePairedDelimiter\expec{\langle}{\rangle}%

\begin{document}

\title{Fokker-Planck treatment of nonlinearities in the dispersive coupling of an ion and an optical cavity}

\author{Alan Kahan}
\affiliation{Instituto de F\'{i}sica Enrique Gaviola, CONICET and Universidad Nacional de C\'{o}rdoba,
Ciudad Universitaria, X5016LAE, C\'{o}rdoba, Argentina}

\author{Leonardo Ermann}
\affiliation{Departamento de F\'isica Te\'orica, GIyA, Comisi\'on Nacional de Energ\'ia At\'omica, Buenos Aires, Argentina}
\affiliation{Escuela de Ciencia y Tecnología, Universidad Nacional de San Martín (UNSAM), San Martín, Argentina}
\affiliation{CONICET, Godoy Cruz 2290 (C1425FQB) CABA, Argentina}

\author{Marcos Saraceno}
\affiliation{Departamento de F\'isica Te\'orica, GIyA, Comisi\'on Nacional de Energ\'ia At\'omica, Buenos Aires, Argentina}
\affiliation{Escuela de Ciencia y Tecnología, Universidad Nacional de San Martín (UNSAM), San Martín, Argentina}

\author{Cecilia Cormick}
\affiliation{Instituto de F\'{i}sica Enrique Gaviola, CONICET and Universidad Nacional de C\'{o}rdoba,
Ciudad Universitaria, X5016LAE, C\'{o}rdoba, Argentina}

\date{March 29, 2023}

\begin{abstract}
  We complement previous studies of an ion coupled with an optical cavity in the dispersive regime, for a model which exhibits bistability of different configurations in the semiclassical description. Our approach is based on a truncated evolution in phase space and is intended to explore an especially interesting parameter region where the fully quantum-mechanical solution becomes hard but the crudest semiclassical approach fails to capture essential phenomena. We compare the results of our techniques with the ones from numerical diagonalization of the quantum evolution and find that although the treatment leads to a smoothening and a slight shift of the transitions in the system, it still provides a clear improvement over localized semiclassical approximations.
\end{abstract}

\maketitle

\section{Introduction}

Along the road toward quantum information processing, several platforms have proven to be promising for the implementation of different kinds of quantum simulations \cite{Altman_2021}. Trapped ions are one prominent example of a system whose level of controllability allows for the simulation of various kinds of physical models of interest, including tunable spin Hamiltonians \cite{Porras_2004, Monroe_2021}, continuous-variable systems \cite{Bermudez_2013, Cormick_PRA_2016}, and hybrid models involving both discrete and continuous degrees of freedom \cite{Mezzacapo_2012, Lemmer_2018}.

One step forward in the manipulation of trapped ions was given by the integration of optical potentials, cavities, and lattices \cite{Pachos_2002}. These provide mechanisms to apply controlled displacements in phase space \cite{Schmiegelow_2016}, pin or tailor the crystal structure \cite{Linnet_2012}, modify the vibrational spectrum \cite{Pruttivarasin2011}, introduce infinite-range interactions \cite{Ramette_2022}, realize different quantum phases \cite{Schmied_2008}, cool the ion motion \cite{Fogarty_2016}, or perform non-invasive measurements through the optical spectrum \cite{cormick2012structural}. Optical potentials were also proposed as a means to perform experiments involving both neutral and charged particles \cite{Cormick_2011}.

The interplay of an optical potential and the mutual Coulomb repulsion between ions can be used to realize a version of the Frenkel-Kontorova model \cite{Garcia-Mata2007}, exhibiting a sliding-pinned transition \cite{bylinskii2015, bylinskii2016observation}. If the optical potential is due to the interaction with a cavity field, for large enough dispersive cooperativities one can find significant back-action of the ion positions on the optical field \cite{fogarty2015}. This effect leads to modifications in the sliding-pinned transition, that can become of first order. A similar change was studied for the linear-zigzag transition of an ion chain \cite{cormick2012structural}.

Due to the computational complexity of the models, theoretical treatments usually rely on semiclassical descriptions which cannot be applied in the vicinity of critical points or when the system is tunneling between different semiclassical configurations. A fully quantum treatment of a simplified system involving only one motional degree of freedom and one cavity mode was presented in \cite{Kahan_2021}. This article found qualitative agreement with several semiclassical predictions but did not detect a closing of the spectral gap signaling classical bistability, which was attributed to the limitations in system size imposed by the numerical diagonalization technique.

In this work, we extend that study by considering an alternative treatment based on a truncated Wigner approximation (TWA), a phase-space method that leads to a Fokker-Planck equation that can be efficiently simulated \cite{Carmichael_1993_book, gardiner_book, Verstraelen_2020, Huber_2021}. We compare the asymptotic state found with this technique with the results of the full diagonalization of the Liouvillian, as well as with the semiclassical description in terms of Gaussian states. This allows us to find the parameter regions where the TWA is more accurate. 

Despite the limitations of the Fokker-Planck approach, we conclude that it can provide a remarkable improvement over localized semiclassical descriptions as in \cite{fogarty2015, cormick2012structural}. For instance, the method we use properly describes relaxation to a stationary state in regimes in which the linearized treatment is not applicable. Furthermore, our nonlinear procedure correctly predicts a smooth crossover between configurations, which cannot be captured by the localized approximation.
This is achieved in a computationally inexpensive manner while recovering the same results as the localized Gaussian approximations in the appropriate limits. The method is conceptually simple and versatile, so it can be applied to problems with more degrees of freedom, and to other quantum-mechanical systems where non-linearities play a relevant role and for which a semiclassical phase-space method is suitable \cite{vicentini2018, lescanne2019,hwang2018,dettmer2001,zhang2021}.

The article is organized as follows: In Sec. \ref{sec:model} we describe the system we consider and the equations governing its evolution. A standard localized semiclassical approximation applied to this model is explained in Sec. \ref{sec:standard semiclassical}, whereas the TWA is presented in Sec. \ref{sec:TWA}. This method is compared with full diagonalization to characterize the asymptotic state of the system in Sec. \ref{sec:meanvalues}. In Sec.~\ref{sec:relaxation} we study the predictions for the relaxation rate to the asymptotic state using phase-space methods. Finally, in Sec.~\ref{sec:conclusions} we present our main conclusions. More technical details are provided in four appendixes.

\section{The optomechanical system}
\label{sec:model}

\begin{figure}[htb]
 \begin{center}
 \subfigure[]{\includegraphics[width=0.55\columnwidth]{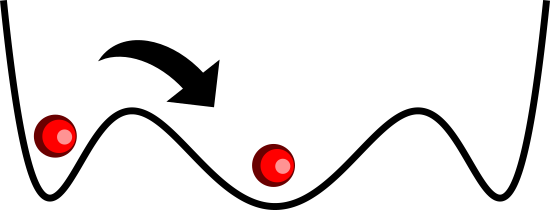}}
 \subfigure[]{\includegraphics[width=0.55\columnwidth]{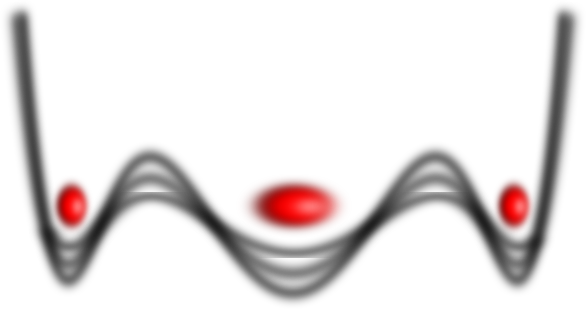}}
 \end{center}
 \caption{Sketch of an atom in a trap, dispersively interacting with a cavity in a classically multistable regime. The trap confines the atom in its center, whereas the cavity field pushes the atom towards field minima located at the sides; the strength of the cavity field depends in turn on the atomic position. (a) Semiclassical description with a fixed effective potential. (b) Sketch of a different semiclassical picture allowing for statistical superpositions in the state of cavity field and atomic position.
 \label{fig:sketch}}
\end{figure}

The model we study is the same as in \cite{Kahan_2021}: it involves a single mode of an optical cavity pumped by a laser and integrated into a harmonic ion trap, within the dispersive regime. We are interested in the situation in which the trap and the optical forces on the ion compete giving rise to classical multistability as illustrated in Fig.~\ref{fig:sketch}. In this section, we provide the main equations for the system evolution. The time dependence of the driving Hamiltonian is eliminated by using the frame rotating with the laser, and thus the total Hamiltonian is given by
\begin{equation}
  H_{S}=H_{\rm cav}+H_{\rm ion}+H_{\text{int}}  \,.
\end{equation}

Here,
\begin{equation}
  H_{\rm cav}=-\hbar \Delta_c a^\dagger a + i \hbar \eta  \left(a^\dagger - a  \right)
\end{equation}

\noindent is the Hamiltonian for the cavity field, with $\Delta_c=\omega_l - \omega_c$ the detuning between the laser pump and the cavity mode, $\eta$ a pumping strength proportional 
to the amplitude of the laser, and $a$ ($a^\dagger$) the annihilation (creation) operator for the cavity field.

The ion's motional degrees of freedom in absence of coupling with the cavity are governed by the Hamiltonian:
\begin{equation}
  H_{\rm ion} = \frac{p^2}{2m} + V_{\rm trap} (x)
\end{equation}
\noindent where for simplicity we consider only one dimension, with the trap potential in harmonic approximation:
\begin{equation}
  V_{\rm trap} (x) = \frac{m\omega^2}{2} x^2 \,.
\end{equation}
If more ions are included in the system, the Coulomb potential must also be taken into account.

The cavity field is coupled to an electronic transition of the ion, with a coupling frequency ${\Omega}(x)$ that depends on $x$ due to the spatial profile of the cavity mode. When cavity and electronic transition are far detuned one can eliminate the internal degrees of freedom of the ion obtaining the standard optomechanical interaction \cite{ritsch2013}:
\begin{equation}\label{Hsys}
  H_{\text{int}}= \hbar \frac{\Omega^2(x)}{\Delta_0} a^\dagger a\,.
\end{equation}
\noindent Here we take $|\Delta_0| \gg |\Delta_c|$, $\Omega$, with $\Delta_0=\omega_l - \omega_0$ the detuning of the pump with respect to the atomic transition. 

The resulting coupling can be interpreted as a frequency shift of the cavity depending on the ion position, which leads to an effective detuning $\Delta_{\rm eff}$ given by: 
\begin{equation}
  \Delta_{\text{eff}} (x) = \Delta_c - \frac{\Omega^2(x)}{\Delta_0} \,.
  \label{eq:Delta}
\end{equation}

\noindent We assume that $\Delta_0$ is sufficiently large to neglect spontaneous emission. For definiteness, we consider a blue-detuned laser, $\Delta_0>0$, so that the ion is attracted to the minima of the optical intensity field.

Finally, the cavity field loses photons at rate $2\kappa$ as described by the master equation:
\begin{equation}
 \mathcal{L}_\kappa = \kappa (2 a \rho a^\dagger - \{a^\dagger a, \rho\} )
\end{equation}
with the curly bracket denoting an anticommutator. This dissipative element is responsible for the relaxation of the system. The above equation describes an environment that cannot introduce photons into the system, a standard assumption in the optical range. We note that so far we have not included any direct dissipation on the ion motion; the existence of a stationary state for the whole system in this model is a consequence of the coupling between ion and cavity.

It is standard to define the dispersive cooperativity $C=U_0/\kappa$, where $U_0=\Omega_0^2/\Delta_0$ with $\Omega_0$ a characteristic value for $\Omega$. The parameter $C$ plays a key role in the dynamics, since it quantifies the impact of the ion position on the cavity field. A small $C$ leads to the usual potential of a dipole trap, whereas $C\gtrsim1$ corresponds to a deformable potential which can also provide cavity-mediated interactions between different atoms.

We focus on a Hamiltonian preserving spatial parity of the ion, which allows one to observe a sharp symmetry-breaking transition in the semiclassical treatment. Indeed, within this approximation, one finds a symmetry-broken state in the strong driving regime, so that the ion localizes close to one of the minima of intracavity field intensity. On the contrary, for weak driving, the equilibrium configuration is such that the probability distribution of the ion is symmetric and located around the trap center. 

In particular, we consider an intensity profile given by: 
\begin{equation}
{\Omega}^2(x) = \Omega_0^2 \,  [ (x/x_{\rm eq})^2 - 1 ]^2\,.
\label{eq:intensity profile}
\end{equation}
Although this is not experimentally realistic, only the central region is relevant for our purposes, and in this region our choice is representative of the behavior of a periodic potential superposed with the harmonic trap. The form in Eq.~(\ref{eq:intensity profile}) leads to simple classical equilibrium positions in the two relevant limits: they are located at $x=0$ for weak pumping and $x=\pm x_\text{eq}$ for infinite pump strength. Another desirable feature is that the optical depth per photon associated with the barrier between minima is characterized by $U_0$. Furthermore, the quartic potential simplifies some numerical calculations and allows one to derive analytical results for the classical equilibrium positions \cite{Kahan_2021}.

\section{Localized semiclassical approximation} 
\label{sec:standard semiclassical}

In this section, we shortly review the standard semiclassical description of this system. For more details, we refer the reader to \cite{cormick2013, Kahan_2021}. We consider the evolution in terms of equations of motion of the Heisenberg-Langevin type. The system dynamics are described by small fluctuations around a classical equilibrium configuration, using the expansion $O = \overline{O} + \delta O$ for the operators $x$, $p$ of the ion and for the field quadratures. Here $\delta O$ represents fluctuations with vanishing mean value, and $\overline{O}=\expec{O}$ is the expectation value of the operator, but we use overline for clarity in the notation.

The classical equilibrium values are the solutions satisfying $\dot a = \dot x = \dot p = 0$.
This immediately leads to $\overline p=0$, whereas the field quadratures are determined by
\begin{equation}\label{eq:aeq}
  \overline a = \frac{\eta}{\kappa -i \Delta_{\rm eff}(\overline x)}
\end{equation}
\noindent and the equilibrium positions of the ion must obey:
\begin{equation}\label{eq:xeq}
  \frac{d~}{d \overline x} \left[ V_{\rm eff}(\overline x) + V_{\rm trap} (\overline x) \right] = 0,
\end{equation}
\noindent with $V_{\rm eff}$ the optical effective potential \cite{maunz2001}
\begin{equation}\label{eq:Veff}
  V_{\rm eff}(\overline x)=-\frac{\abs{\eta^2}}{\kappa}\arctan{\left(\frac{\Delta_{\rm eff}(\overline x)}{\kappa}\right)} \,.
\end{equation}
Although Eq.~(\ref{eq:xeq}) only imposes a zero derivative of the total effective potential, stability considerations show that the classical equilibrium positions must correspond to minima of the total effective potential. Further stability requirements are explained in \cite{cormick2013}.

The semiclassical description to lowest order predicts a stable
equilibrium position at $\overline x=0$ for weak laser pumping, whereas for large pumping strength, the stable equilibrium positions approach $\pm x_{\rm eq}$.
The transition between both kinds of solutions can be continuous or discontinuous depending on the cooperativity $C$ and the laser detuning \cite{Kahan_2021}. 

The focus of this work is the treatment of the regime of semiclassical multistability. Here we will refer to bistability focusing on the transition between the configuration with the ion located at the center or at the sides, regardless of the difference between left and right. We note that the system evolution we consider does not couple states with odd and even parity in the ion variables. We expect the relevant quantum states to describe the transition to be those in the even parity subspace.

The semiclassical approach which was the basis of the studies in \cite{cormick2012structural, cormick2013, fogarty2015, Fogarty_2016} also includes a linear treatment of the fluctuations, truncating the Heisenberg-Langevin equations to first order in the displacements from the mean values. In the following, we refer to this as the ``localized Gaussian approximation''. For the particular model we consider here, we provide the equations in Appendix \ref{sec:Gaussian} and refer the reader to \cite{cormick2013} for more details.
In this way, one can find an asymptotic Gaussian state for the system if the parameter regime corresponds to a stable configuration. Stability considerations for the fluctuations have been discussed in detail in \cite{cormick2013}. In particular, cavity cooling of the ion requires a negative effective detuning $\Delta_{\rm eff}$ and also a non-negligible coupling between the cavity and the motional fluctuations. 

In absence of direct dissipation on the ion, the asymptotic state is not always well defined. In our model, this happens when the ion is located exactly at the center, since to lowest order the ion-cavity coupling vanishes in this case \cite{cormick2013}. This Gaussian treatment also predicts very poor cavity cooling when the ion position approaches $\pm x_{\rm eq}$, because the coupling also vanishes at those points. Furthermore, when $\Delta_c=0$, $\Delta_{\rm eff}$ approaches zero when the ion localizes close to $\pm x_{\rm eq}$. In general, this is associated with more excited asymptotic motional states \cite{Fogarty_2016}. We note, nevertheless, that it is straightforward to add to the model dissipative channels acting directly on the ion motion reproducing experimental sources of noise and cooling that can also be relevant for the determination of the asymptotic state.

\section{One further step in the semiclassical approach}
\label{sec:TWA}

The semiclassical treatment described in the previous Section is expected to become more accurate in the limit with a large number of photons and with $x_{\rm eq} \gg x_\omega=\sqrt{\hbar/(m\omega)}$, i.e. when both degrees of freedom have large effective system sizes so that quantum fluctuations can be considered comparably small. Nevertheless, even under these assumptions, the description breaks down close to the instability points due to the large position fluctuations. Furthermore, this procedure cannot describe solutions which are statistical mixtures of states located around different semiclassical solutions, and we expect this kind of situation to be generic within the regime of semiclassical bistability. 

In the following, we study an extension of this treatment in which all semiclassical configurations can coexist and need not be approximated by Gaussian states. For this, we work with a phase-space representation without fully linearizing the potential, in a similar spirit as \cite{vicentini2018, verstraelen2018,Verstraelen_2020,Huber_2021}. In order to make calculations feasible, we do perform a truncation of the high-order derivatives, and we compare the results with the fully quantum-mechanical treatment to characterize the reliability of the method.

It is convenient to work with dimensionless variables to obtain a uniform notation for the phase-space representation of both the cavity field and the ion motional state. For the cavity, we make the usual choice:
\begin{eqnarray}
 q_1 &=& \frac{a+a^\dagger}{\sqrt{2}}\\
 p_1 &=& -i\, \frac{a-a^\dagger}{\sqrt{2}}
\end{eqnarray}
whereas for the ion motion we eliminate the dimensions with the transformation:
\begin{eqnarray}
 x &\to& q_2\, x_\omega\\
 p &\to& p_2\, p_\omega
\end{eqnarray}
with $x_\omega=\sqrt{\hbar/(m\omega)}$ and $p_\omega=\sqrt{\hbar m \omega}$ determined by the ion's mass and the trap frequency, so that $q_2$ and $p_2$ are dimensionless.
The Wigner representation for the two-mode state $\rho$ is then defined as \cite{Schleich_book}:
\begin{equation}
 W_\rho (\vec q, \vec p) = \frac{1}{(2\pi)^2} \int du_1 du_2 e^{-i\vec p \cdot \vec u} \bra{\vec q+ \vec u/2} \rho \ket{\vec q - \vec u /2}
\end{equation}
where we use the shorthand $\vec q = (q_1,q_2)$,  $\vec p = (p_1,p_2)$.

The equation for the time evolution of the Wigner function can be found following standard procedures \cite{gardiner_book}. This equation is non-linear in the system variables and also contains high-order derivatives. We consider only derivatives up to second order and discard the rest. The resulting equation is:
\begin{widetext}
 \begin{multline}
  \frac{\partial W}{\partial t} \simeq -\sqrt{2} \eta \frac{\partial W}{\partial q_1} 
      + \omega \left( q_2 \frac{\partial W}{\partial p_2} - p_2 \frac{\partial W}{\partial q_2} \right) - \frac{1}{2} \frac{d \Delta_{\rm eff}}{dq_2} (q_1^2+p_1^2-1) \frac{\partial W}{\partial p_2} - \Delta_{\rm eff} (q_2) \left( q_1 \frac{\partial W}{\partial p_1}  - p_1 \frac{\partial W}{\partial q_1} \right) \\
      + \kappa \left[\frac{\partial ~}{\partial q_1} (q_1 W) + \frac{\partial ~}{\partial p_1} (p_1 W) + \frac{1}{2} \frac{\partial^2 W}{\partial q_1^2} + \frac{1}{2} \frac{\partial^2 W}{\partial p_1^2} \right]
\,.
 \label{eq:TWA}
 \end{multline}
\end{widetext}

The equation can be recast in terms of classical Poisson brackets for the terms $H_{\rm ion}$,  $H_{\rm cav}$, $H_{\rm int}$, with $\mathcal{L}_\kappa$ providing the extra dissipative and diffusive terms contained in the second line of Eq.~\eqref{eq:TWA}. It is clear then that the approximation yields classical evolution -- including the nonlinearities of the intensity profile -- for the Wigner density modified by diffusive and dissipative terms proportional to $\kappa$.

It has been noted that, when a localized semiclassical solution is assumed, then this equation must also be linearized through an expansion of the Hamiltonian around the equilibrium position. This is because the discarded higher-order derivatives have contributions that are of the same order as the Hamiltonian terms of order higher than two around the classical equilibrium \cite{Carmichael_1993_book}. This argument, however, cannot be applied if one wishes to consider situations where there is more than one classical equilibrium configuration.

One can, on the other hand, argue that discarding the higher-order derivatives is justified in view of the behavior of the different terms in the equation for the evolution of the Wigner function when a semiclassical limit is taken. When dimensional variables are used, this limit can be thought of as arising when $\hbar \to 0$. Since we take dimensionless variables, the equivalent limit can be obtained by rescaling the relevant quantities by a factor $f>1$ and then considering $f\to\infty$. In our system, the semiclassical rescaling would be:
\begin{align}
 q_j \to f q_j, \quad p_j \to f p_j, \quad \eta \to f \eta
 \label{eq:scaling}
\end{align}
whereas, on the other hand, the quantities $\kappa$, $\Delta_c$, and $\omega$ must be left constant. For the interaction term, we keep the order of magnitude of the effective detuning constant, while the optical equilibrium positions are changed by the scale factor $f$. This means that while we perform the transformations \eqref{eq:scaling}, $\Delta_{\rm eff}(q_2)$ is left unchanged.

When this scaling transformation is applied, all terms in Eq.~\eqref{eq:TWA} remain unchanged except for the noise terms which are proportional to $\kappa$ and contain second derivatives with respect to the field quadratures. These terms, indeed, decrease by a factor $1/f^2$, which is consistent with the idea that quantum fluctuations become comparatively smaller as the system is taken towards the semiclassical limit. The terms having higher derivatives and which we have discarded in Eq.~\eqref{eq:TWA} are accompanied by prefactors $1/f^4$ or smaller. 

\begin{figure*}[t]
  \includegraphics[width=1\textwidth]{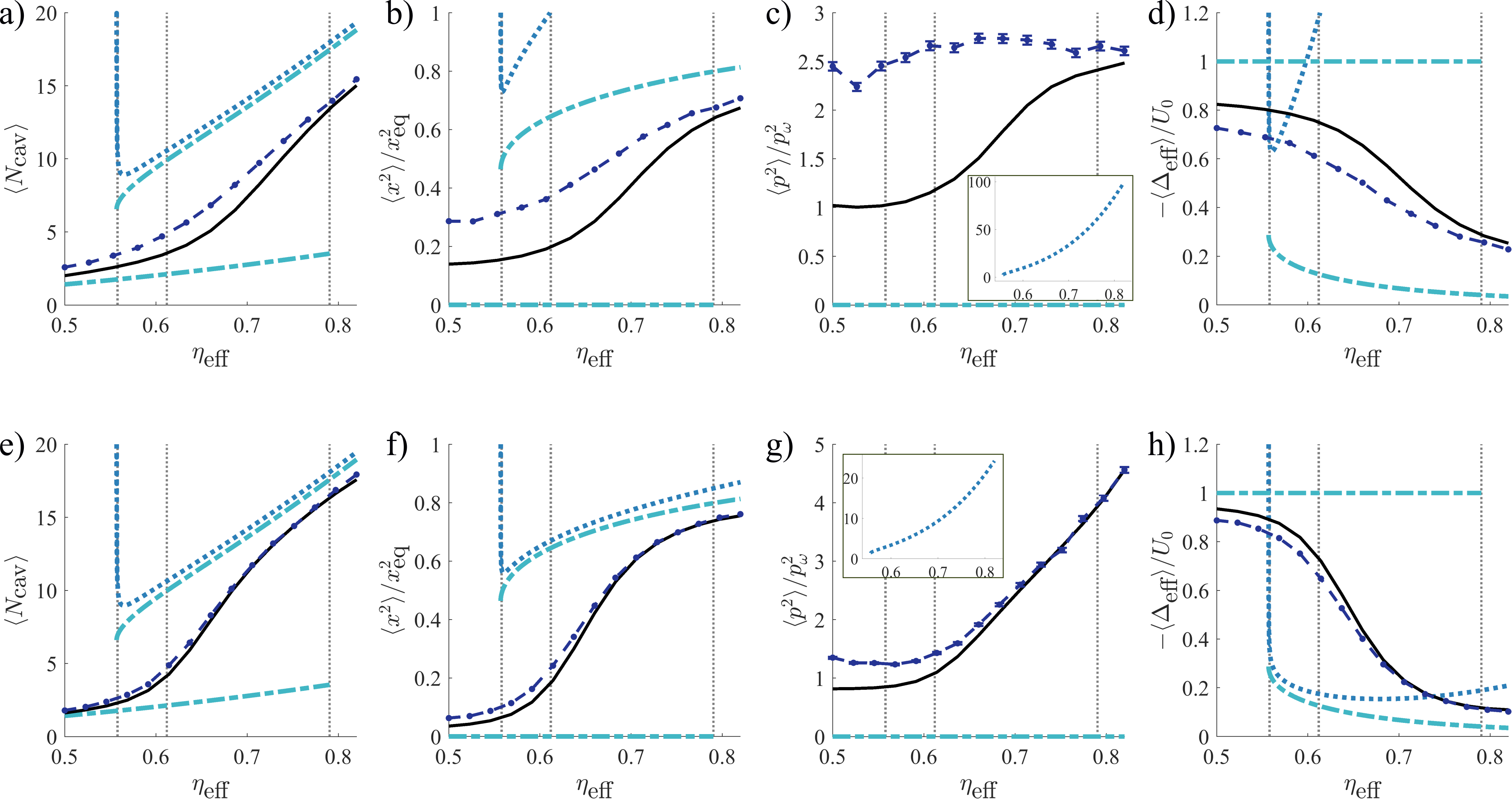}
  \caption{Mean values characterizing the steady state as a function of the effective pump strength in the different treatments considered: exact numerical diagonalization (solid black), TWA (circles joined by dashed dark blue line), semiclassical prediction to lowest order (dash-dotted light blue), localized Gaussian approximation (dotted blue, only on the regime where the ion is at the sides). In all cases, the cooperativity and cavity detuning are fixed to $C=2$ and $\Delta_c=0$ respectively. The spatial scale given by $x_{\rm eq}/x_\omega$ is chosen as 3 (top) and 7 (bottom). The value of $\kappa/\omega$ is varied in order to obtain the same lowest-order semiclassical predictions for the two cases. The vertical dashed lines on the left and right surround the region of classical bistability, whereas the dashed line between them indicates the change in global minimum of the effective potential. The insets in (c) and (g) show the behavior of the Gaussian approximation, which is out of the scale of the main plots. In subplots (b) and (d) we cut the vertical axis leaving out part of the dotted curve to facilitate the comparison between TWA and diagonalization.
  \label{fig:FP_arnoldi_xeq}
  }
\end{figure*}

The truncation given in Eq.~\eqref{eq:TWA} leads exactly to a Fokker-Planck equation, with cavity losses introducing drift and diffusion terms. As explained in Appendix \ref{sec:stochastic}, this evolution can be mapped to a stochastic process with straightforward and economical numerical implementation  \cite{Carmichael_1993_book}. However, this mapping implies that one needs to average over many realizations.

The resulting phase-space distribution is always positive. Although positivity of the Wigner function is one standard criterion for classicality, states with positive Wigner functions may nevertheless exhibit entanglement or be useful for quantum metrological tasks. In any case, the procedure we follow does not lead to ``more quantum'' states than the ones obtained with the localized semiclassical method as in \cite{cormick2012structural, fogarty2015}. It represents an improvement in the sense that it makes it possible to describe statistical superpositions of different localized semiclassical configurations in a simple manner and without restricting the dynamics to Gaussian states. We stress that this is not the same as the improved semiclassical approximations that include interference between different paths \cite{brack_2018, reichl_1992}.

\section{Numerical results: Characterization of the asymptotic state}
\label{sec:meanvalues}

In the following, we compare the predictions for the steady state obtained from the Fokker-Planck (FP) implementation described in the previous section with those from numerical diagonalization of the Liouvillian as explained in \cite{Kahan_2021}. We are interested in the regime of moderate cooperativities, such that the classical model can exhibit multistability but low enough that they are close to experimentally achievable values. As mentioned before, two different semiclassical limits can be defined. The first one is expected to appear as the intracavity photon number is increased, while the other one is reached when the optical potential wells are separated by length scales much larger than the natural scale of the harmonic trap. The truncated Wigner approximation (TWA) as in Eq.~\eqref{eq:TWA} is expected to be valid when both semiclassical limits are justified. 

We characterize the pump strength by a dimensionless effective laser amplitude
\begin{equation}
\eta_{\rm eff} = \frac{\eta}{\sqrt{\kappa \omega}} \frac{x_\omega}{x_{\rm eq}}
\end{equation}
which is the relevant parameter combination according to the semiclassical study in \cite{Kahan_2021}. For definiteness, we fix $C=2$ and $\Delta_c=0$ in all plots. While there is nothing special about $C=2$, choosing the laser frequency to be resonant with the bare cavity frequency highlights some of the advantages of the improved semiclassical treatment we use. This is because, as discussed in Sec.~\ref{sec:standard semiclassical}, with $\Delta_c=0$ the standard Gaussian treatment predicts highly excited motional asymptotic states for large pumping.

Experimentally achievable values for the different parameters vary according to the realization considered. 
Typical trap frequencies can be in the range $\omega=2 \pi \times$ (47--364) kHz
\cite{bylinskii2015,Linnet_2012,schneider2010}, or even $\omega=2 \pi \times 2.7$ MHz \cite{steiner2013}. Typical values of the single-photon coupling rate, which sets the scale for $\Omega_0$, are $2 \pi \times$ (0.53--6) MHz \cite{thompson1992,steiner2013,herskind2009}.
The value of $x_{\textrm{eq}}$ also depends on the implementation in mind. One can think of our fourth-order potential as a qualitative approximation of a sinusoidal optical intensity around the ion trap center. 
Thus, the dimensionless equilibrium position corresponds to $x_{\textrm{eq}}/x_{\omega}=\lambda/4 x_{\omega}$, where $\lambda$ is the wavelength of the field mode. As example systems we consider Yb atoms with laser light of 369 nm \cite{bylinskii2015} and Ca atoms with 405 nm \cite{Linnet_2012}, and then $x_{\textrm{eq}}/x_{\omega} \approx$ 2--7, up to $x_{\textrm{eq}}/x_{\omega} \approx 20$ for a high motional frequency $\omega$ of a few MHz. We note that one could also think of alternative models where the motion along the cavity axis is irrelevant or pinned, and then the optical potential would correspond to the transverse optical profile. Experimental values of the cavity decay rate are very variable and generally lie within the range $\kappa=2\pi \times$ 68 kHz--320 MHz \cite{lee2018,herskind2009,steiner2013,meraner2020}.

To compare the Fokker-Planck equation and the numerical diagonalization, we focus on a few representative expectation values which were previously studied in \cite{Kahan_2021}: the number of photons, the squared position operator of the ion, its kinetic energy, and the effective cavity detuning. These quantities are plotted in Fig.~\ref{fig:FP_arnoldi_xeq}. In each subplot, the  solid black line corresponds to numerical diagonalization and the dashed dark blue curve is the one obtained from TWA. For comparison we also plot the semiclassical approximation to lowest order corresponding to Eqs.~(\ref{eq:aeq})--(\ref{eq:Veff}) (dashed-dotted light blue) and the localized Gaussian approximation explained in Appendix~\ref{sec:Gaussian} (dotted blue).

We remind the reader that the lowest-order semiclassical approximation exhibits bistable behavior, which is why there are two such solutions for intermediate values of $\eta_{\rm eff}$. In contrast, the localized Gaussian description (which consequently shows bistability) is only displayed for the solution with the ion at the sides. This is because we are considering a case with no direct dissipation on the ion, so that the linearized treatment does not lead to a unique steady state when the ion is located at the center (see Appendix \ref{sec:Gaussian}). We also note that the equations corresponding to the localized Gaussian approximation become unstable at the point where the semiclassical solutions with $\overline x\neq 0$ disappear as the pumping strength is decreased. Regarding the TWA, the sampling error in the mean values obtained is below the marker size, with the exception of $\langle p^2 \rangle/p_\omega^2$, for which error bars are shown.

The top and bottom rows of Fig.~\ref{fig:FP_arnoldi_xeq} correspond to $x_{\rm eq}/x_\omega=3$ and 7 respectively, with $\kappa$ and $U_0$ chosen so that the cooperativity stays fixed and the lowest-order semiclassical solutions are the same in both cases. This choice allows us to observe how the accuracy of the TWA varies with the spatial scale for the ion while the scale of the photon field is kept fixed. 
As expected, the TWA provides more reliable results in the regime  with a larger spatial scale of the optical potential (bottom row) and with a larger number of photons (corresponding to the right part of each figure).

\begin{figure*}[ht]
  \includegraphics[width=1\textwidth]{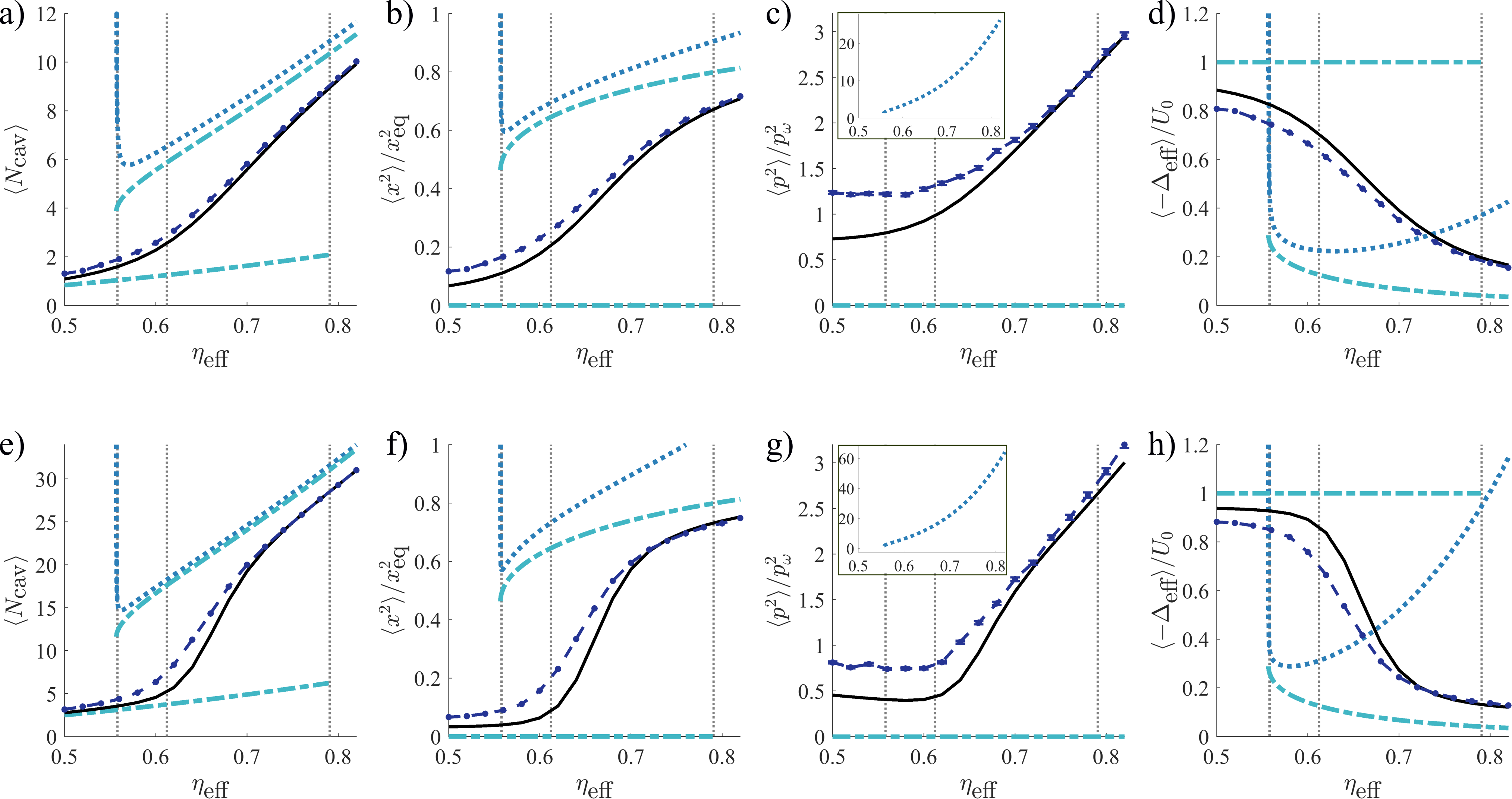}
  \caption{
  Mean values characterizing the steady state as a function of the effective pump strength in the different treatments considered. The color code is the same as in the previous figure. In all cases, the cooperativity and cavity detuning are fixed to $C=2$ and $\Delta_c=0$ respectively, while the spatial scale is chosen as $x_{\rm eq}/x_\omega=5$. The values of $\kappa/\omega$ are $1.5$ (top row) and $0.5$ (bottom row). The vertical dashed lines on the left and right surround the region of classical bistability, whereas the dashed line between them indicates the change in global minimum of the effective potential.  The insets in (c) and (g) show the behavior of the Gaussian approximation, which is out of the scale of the main plots. In subplot (f) we cut the vertical axis leaving out part of the dotted curve to facilitate the comparison between TWA and diagonalization.}
  \label{fig:FP_arnoldi_N}
\end{figure*}

Another noticeable point in Fig.~\ref{fig:FP_arnoldi_xeq} is that the FP predictions are less accurate for the ion dispersion in both position and momentum for low pumping. One must keep in mind that even if for low photon numbers the ion motion is mostly determined by the trap potential, in absence of direct dissipation on the ion it is only the cavity that provides cooling of the ion motion. For lower photon numbers the effect of truncating photon fluctuations can become more important both in the determination of the optical potential and in the cooling mechanisms. 

Nevertheless, the TWA outperforms by far the localized Gaussian approximation in the full range studied. When the ion is located at the trap center, the linearized Gaussian approximation predicts a vanishing coupling between ion and cavity, and therefore the asymptotic state of the ion is not well defined in absence of additional dissipation (see Appendix \ref{sec:Gaussian}). The TWA does not suffer from this problem, and the system has a steady state even in this particular case. When the ion is located at the sides, the lowest-order coupling is non-vanishing but, as mentioned in Sec. \ref{sec:standard semiclassical}, the linearization predicts very poor cavity cooling as $\Delta_{\textrm{eff}}(\overline x)$ becomes very small. This leads to the excessive growth of the ion's kinetic energy, as shown in the insets of Fig.~\ref{fig:FP_arnoldi_xeq}-c) and g). In contrast, the TWA correctly accounts for the effect of the ion's spatial spread on the effective detuning and thus predicts kinetic energies much closer to the ones obtained through numerical diagonalization.

In Fig.~\ref{fig:FP_arnoldi_N} we perform a similar comparison but leaving $x_{\rm eq}$ fixed and varying $\kappa/\omega$, $U_0/\omega$ so that the number of photons in the transition region changes (the ratio $C=U_0/\kappa=2$ is kept fixed). 
We show asymptotic mean values for $\kappa/\omega=1.5$ and $0.5$ in the top and bottom rows  of Fig.~\ref{fig:FP_arnoldi_N}, respectively. In all plots, we observe good agreement for strong pumping. As in the previous case, we notice that the TWA predicts larger values of squared position and momentum for the ion at low pumping. However, and in contrast with the behavior in the previous figure, the TWA does not become more accurate for curves with higher mean photon numbers. Indeed, higher photon numbers lead to a more abrupt crossover between configurations, but in the TWA this happens more smoothly than in the exact diagonalization. Within the TWA treatment the transition is also slightly shifted towards weaker pumping. 

\begin{figure}[h!]
  \includegraphics[width=0.6\columnwidth]{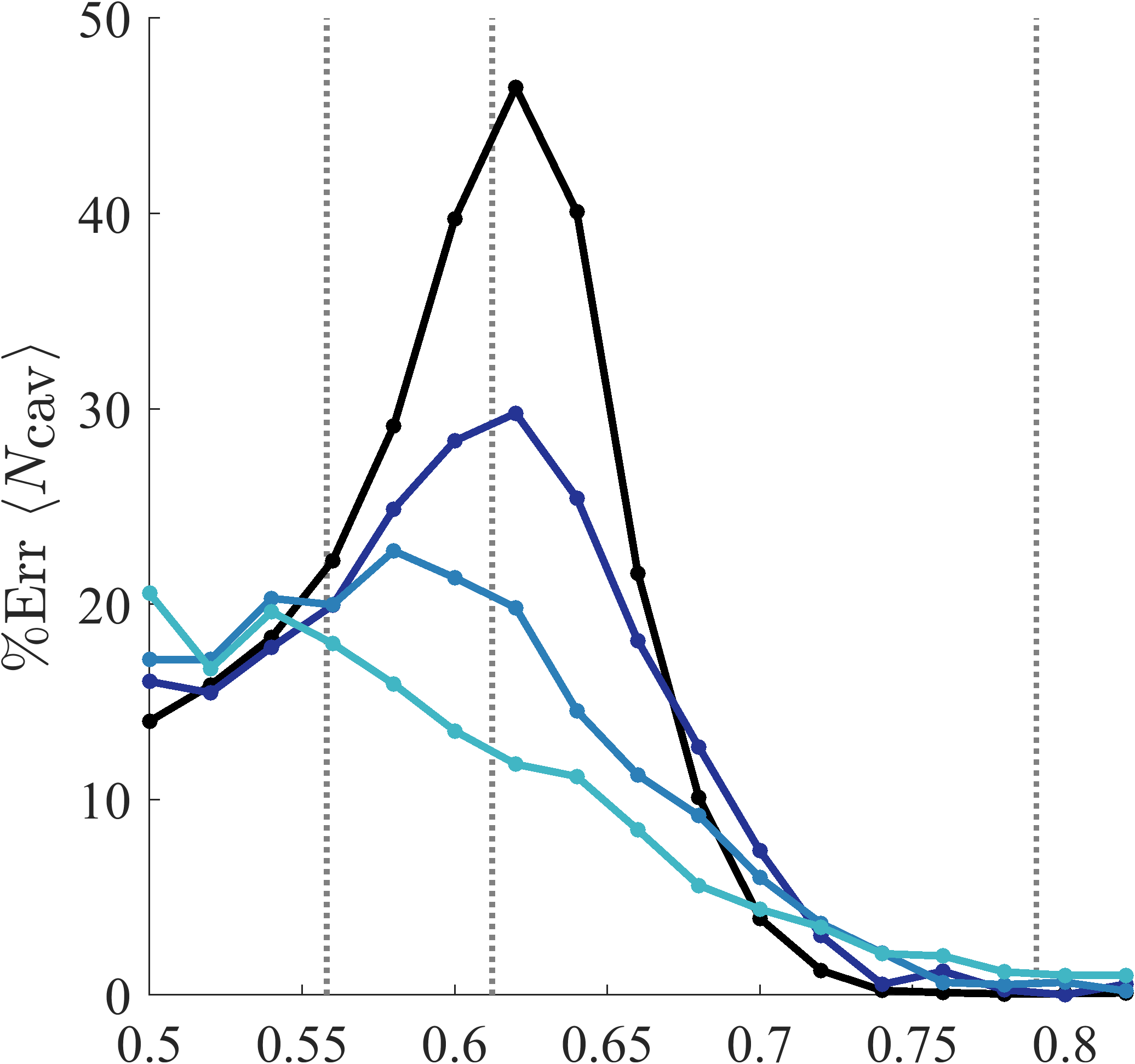}
  \caption{Percent error of the TWA compared to numerical diagonalization. To vary the number of photons we take $\kappa/\omega=0.50, 0.75, 1.00, 1.50$, with darker colors corresponding to lower values of $\kappa$ (higher photon numbers). The remaining parameters have the same values as in Fig.~\ref{fig:FP_arnoldi_N}. Vertical dashed lines have the same meaning as in the previous figures.
  }  \label{fig:err_FP_arnoldi_vs_kappa}
\end{figure}

To better analyze this behavior, in Fig.~\ref{fig:err_FP_arnoldi_vs_kappa} we show the percent error of the results of the TWA for the mean photon number compared to the diagonalization for a larger set of values ($\kappa/\omega=1.5$, $1.0$, $0.75$, $0.5$). We find peaks of the relative error in the transition region, corresponding to the regime of classical bistability. Most importantly, contrary to the naive expectation, the peaks of the percent error become larger for curves with higher mean photon numbers.
A possible explanation for this result is that the dynamics of the system are more strongly non-linear as the transition sharpens, and this effect dominates over the ``more semiclassical'' behavior expected for higher photon numbers. When assessing the reliability of the TWA, one must then keep in mind that larger typical scales do not always lead to better results. Nevertheless, we stress again that for parameter regimes when a fully quantum treatment becomes too costly, the FP method still provides a much better approach than the standard localized Gaussian approximation. Indeed, the latter is totally unable to describe the crossover from one equilibrium configuration to the other. 

Another aspect that deserves attention is the fact that, although the standard semiclassical description correctly predicts the location of the crossover region, the change in the global minimum of the total effective potential not always coincides with what one would identify as the ``transition point'' between configurations. This is a subtle issue: even though the system is dissipative, previous articles may convey the impression that the global minimum of the total effective potential can be identified with a semiclassical ``ground state'', with a privileged role in the dynamics \cite{fogarty2015, Fogarty_2016, Buchheit_2020}. However, it is not obvious that, apart from the condition of having a local minimum, the actual value of the effective potential has any significance in the determination of the asymptotic state.

\begin{figure}[ht]
  \includegraphics[width=1\columnwidth]{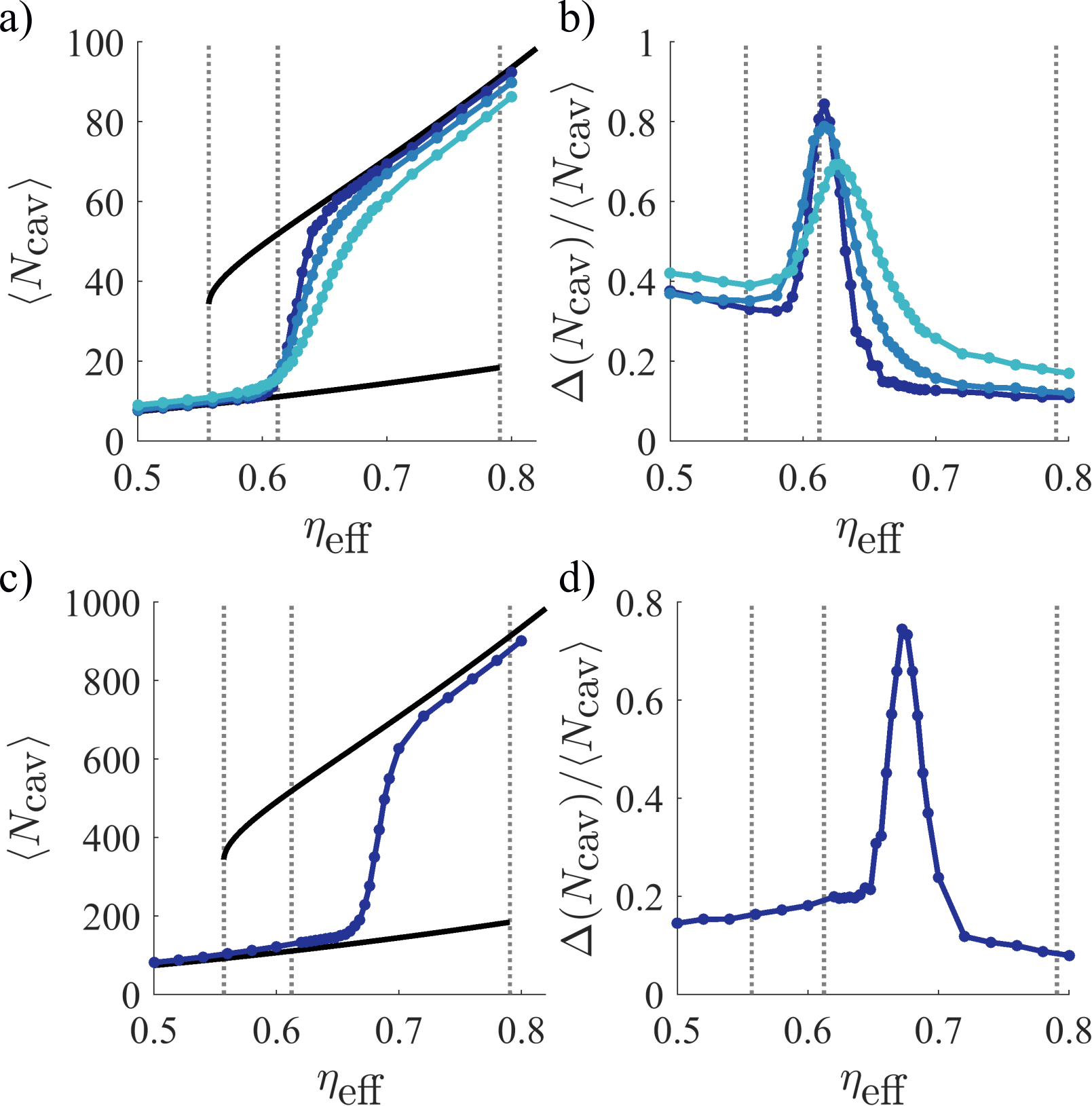}
  \caption{Typical transition markers for moderate (top row) and large (bottom row) mean photon numbers, as a function of the effective pumping strength. Left: mean photon number, right: relative fluctuations in photon number. Parameters used are: $C=2$, $\Delta_c=0$, $\Gamma/\omega=5\times 10^{-4}$. Top row: $\kappa/\omega=0.170,0.551,1.730$ and $x_{\textrm{eq}}/x_\omega=5,9,15$ respectively, with darker blue meaning increasing values of $x_{\textrm{eq}}/x_\omega$. Bottom row: $\kappa/\omega=0.0170$ and $x_{\textrm{eq}}/x_\omega=5$. 
  The dashed lines show the different transitions of the effective potential. The semiclassical prediction to lowest order is shown in black lines.}
  \label{fig:FP_semiclasico}
\end{figure}

This discrepancy is better illustrated in Fig.~\ref{fig:FP_semiclasico}, where we plot the mean photon number and its fluctuations as obtained through the FP method. For this figure, for which the dynamics explores a larger phase space, we included direct dissipation on the ion motion at a rate $\Gamma$ to speed up convergence. More precisely, we introduced extra terms identical to those in the second line of Eq.~\eqref{eq:TWA}, replacing $\kappa$ by $\Gamma$ and $q_1,p_1$ by $q_2,p_2$. The value of $\Gamma$ was chosen very small to reduce the impact of this dissipation on the quantities plotted.

In both the upper and the lower rows of Fig.~\ref{fig:FP_semiclasico}, the lowest-order semiclassical prediction for the photon number is plotted in continuous black lines. In the upper row, the parameters are varied in such a way that the black curve is the same for all cases, and within the transition region photon numbers are around $\langle N_{\rm cav}\rangle\simeq 30$. In this case, as the values of $x_{\rm eq}/x_\omega=5,9,15$ are increased the numerical results approach the semiclassical prediction: the transition becomes sharper and approaches the point of change in global minimum of the effective potential. 

In contrast, the lower row displays the results for much higher photon numbers of $\approx 400$ at the transition but with a smaller spatial scale for the ion, $x_{\rm eq}/x_\omega=5$. In this case, a quite abrupt transition is observed but at a location that clearly differs from the shift in global minimum of the effective potential. We verified that the displacement of the transition point with respect to the semiclassical prediction is not due to the presence of dissipation on the ion motion. Moreover, if we take $\Gamma=0$ in Figs.~\ref{fig:FP_semiclasico}(c) and \ref{fig:FP_semiclasico}(d) the transition moves slightly over larger values of $\eta_{\textrm{eff}}$. Since typical experimental values for $x_{\rm eq}$ are rather far from semiclassical, we conclude that the global minimum of the effective potential is generally not a proper indicator of the transition point.

\section{Numerical results: Relaxation to the asymptotic state}

\label{sec:relaxation}

We now consider the approach to the asymptotic state; in a fully quantum treatment, this state is unique if one restricts to the subspace with even parity for the ion degrees of freedom, or if one introduces noise on the ion motion. Systems with classical multistability are expected to display some kind of metastability in the quantum regime, which manifests as a two-step relaxation process: a fast decay into the metastable manifold, followed by a slow convergence towards the true asymptotic state \cite{rose2016, macieszczak2016, macieszczak2020}.

In a previous study using numerical diagonalization for parameter regimes with $\langle N_{\rm cav}\rangle \simeq 10$ and $x_{\rm eq}=5$, there was no sign of metastability in terms of a closing of the spectral gap \cite{Kahan_2021}. It was conjectured that this was due to the small system size and that metastability signatures required either larger numbers of photons or larger values of $x_{\rm eq}$, which would lead to smaller overlaps between the two kinds of semiclassical equilibrium configurations. 

We thus wish to find methods to estimate the behavior of the relaxation time as a function of the system parameters for ``more semiclassical'' regimes for which fully  quantum treatments are computationally costly. To this aim, we considered two different semiclassical procedures. Our first estimate was based on phase-space overlaps, with very unsatisfactory results, and is explained in Appendix \ref{sec:overlaps}. The second attempt analyzed the time evolution of each trajectory according to the Fokker-Planck equation. We note that none of these estimates were expected to be accurate, but rather to provide the right order of magnitude of the relaxation rates. 

In the following we describe this second procedure, extracting the relaxation times directly from the evolution of the stochastic trajectories according to the Fokker-Planck equation associated with Eq.~\eqref{eq:TWA}. To do this we define two phase-space regions that correspond to each semiclassical configuration: one with the ion at the sides and with a large mean photon number, and another with the ion in the central region and a small mean photon number. Our method to define the regions in phase space is based on ellipses surrounding the peaks in the distributions found from FP evolution, and is explained in detail in Appendix~\ref{sec:delimitation}.

In this form, we simplify the problem by mapping it onto a two-state stochastic system governed by the rate equations:
\begin{equation}
  \begin{lcase}
    \dot p_1 & = -\gamma_1 p_1 + \gamma_2 p_2 \\
    \dot p_2 & =  \: \: \: \, \gamma_1 p_1 - \gamma_2 p_2 \\
  \end{lcase}
\end{equation}
Here, $p_{j}$ represents the probability to find the system in region $j$, whereas $\gamma_j$ is the transition rate out of region $j$, with $j=1,2$. Each state is assimilated with one of the regions, and the equilibration rate towards the steady state is then given by $\gamma_t=\gamma_1+\gamma_2$. We regard this as a valid approximation after a fast relaxation of the initial state has taken place, leaving the dynamics occurring mainly within a metastable manifold over a much slower timescale.

The relaxation rates of a Markov model are closely related to the mean first passage time between different states \cite{okushima2019}, which has been useful to find clusters of metastable states \cite{Kells_2019}. 
We then perform a time evolution of an ensemble of trajectories according to the stochastic process described by the Fokker-Planck equation and relate the mean first passage time $\tau_i$ with the corresponding transition rate as $\tau_i=1/\gamma_i$. Here, we consider the passage time as the time it takes for the trajectory to reach the phase-space region corresponding to the other state for the first time. We note that the results depend on both the definition of the regions and the initial distribution in phase space, so that, again, only qualitatively correct trends should be expected. 

\begin{figure}[ht]
  \includegraphics[width=0.6\columnwidth]{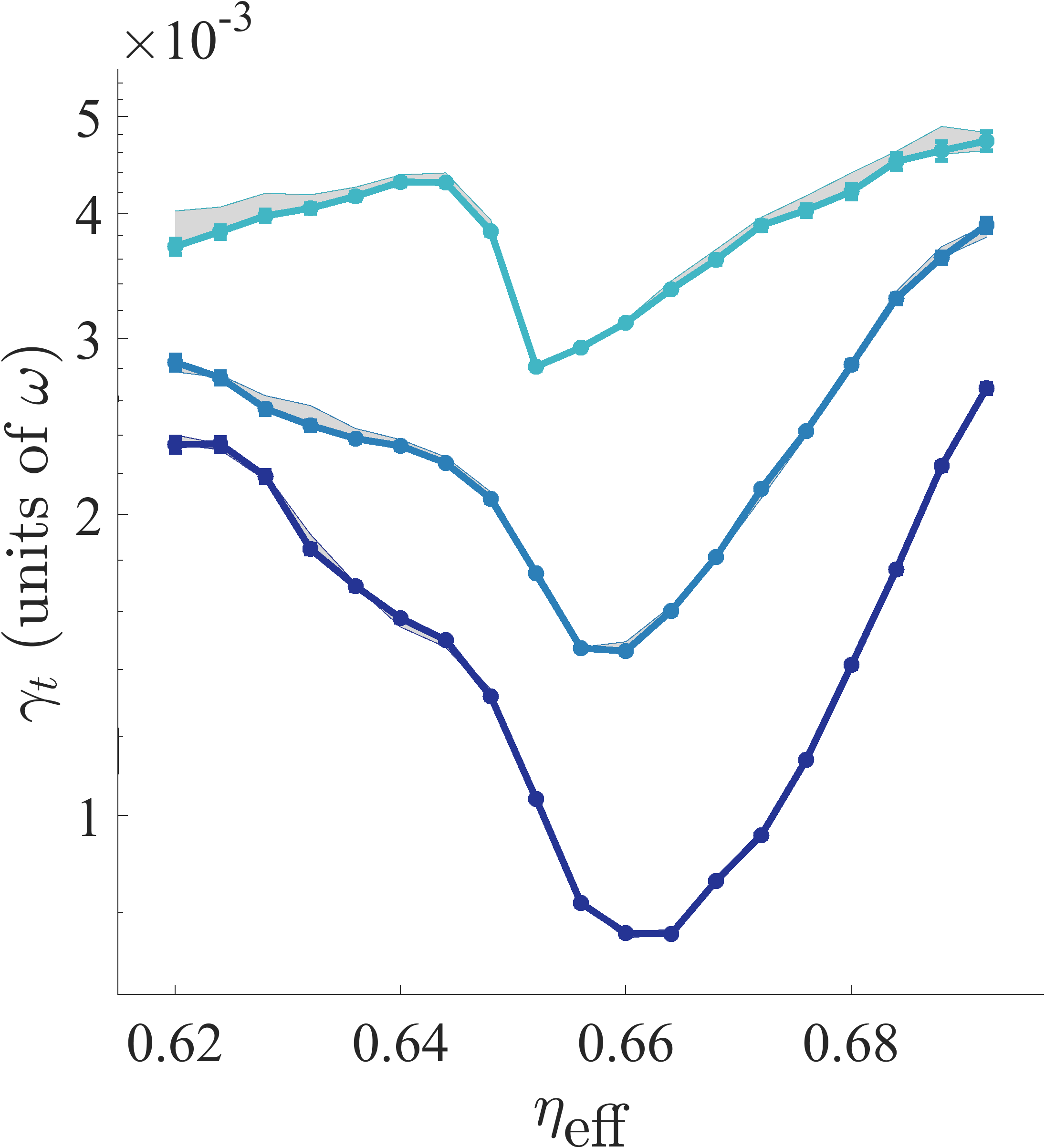}
  \caption{Tunneling rates obtained through the TWA, for $\kappa=0.1\,\omega$, $C=2$, $\Gamma=0$. The spatial scale given by $x_{\rm eq}/x_\omega$ is chosen as 5, 6, 7 for the blue curves, with darker blue for higher $x_{\rm eq}$ (and correspondingly also to higher photon numbers). Notice the logarithmic vertical axis. Error bars (only shown for errors larger than the marker size) correspond to sampling errors, shaded areas indicate how the results change with different choices of the regions (see Appendix~\ref{sec:delimitation} for details).
  \label{fig:FP_tunneling_rates}}
\end{figure}

In Fig.~\ref{fig:FP_tunneling_rates} we show the relaxation rate $\gamma_t$ for a fixed value of $\kappa$ while varying $x_{\rm eq}$, so that larger values of $x_{\rm eq}$ require larger numbers of photons to shift the equilibrium configuration. Darker blue corresponds to ``more semiclassical'' regimes characterized by both larger spatial scales and larger numbers of photons. The shaded areas indicate outcomes for slightly larger or smaller regions (see Appendix~\ref{sec:delimitation}). The results clearly show a slower relaxation as the semiclassical regime is approached, as expected since the phase-space regions corresponding to the different equilibrium configurations become more distant. One can also notice a slower relaxation in the central region of the crossover that could signal a critical point in the truly semiclassical limit. We note that we do not apply this method to the full classically bistable region since the phase-space regions cannot be properly defined in the vicinity of the stability boundaries.

The regime studied in Fig.~\ref{fig:FP_tunneling_rates} is numerically very hard to simulate in a fully quantum treatment. It is then difficult to assess the reliability of these results. As a benchmark, we performed simulations of only one point (the one with the largest error on Fig.~\ref{fig:FP_tunneling_rates}) using 6400 trajectories within the quantum jumps formalism \cite{Dalibard_1992, Carmichael_1993_book}. Our results show good agreement of the mean values, while the relaxation rates found from numerical fits are about two to three times larger for quantum jumps than for Fokker-Planck. We thus conclude that while the rates found from TWA are, as anticipated, not accurate, they still provide qualitative agreement with the expected behavior and the right order of magnitude.

\section{Conclusions and outlook}
\label{sec:conclusions}

We have examined the reliability of extended semiclassical methods for the optomechanical system composed of one atom dispersively coupled with a pumped cavity. We have focused on a regime with classical multistability, which is the one conceptually most interesting and numerically most challenging since Gaussian approximations perform poorly in this parameter region \cite{Kahan_2021}.

Our study is based on a truncated Wigner approximation where derivatives of order higher than two are discarded, but nonlinearities in the system variables are kept. We then write the evolution in terms of a Fokker-Planck equation which can be simulated with very moderate resources. Keeping some higher-order terms makes it possible to describe a system with more than one semiclassical equilibrium configuration, and without restricting to Gaussian states. This proves to be a substantial improvement over the local Gaussian approximation.

We presented a comparison of this procedure with the numerical diagonalization of the quantum evolution superoperator, considering the predictions of the two methods for several quantities characterizing the asymptotic state. The Fokker-Planck approximation generally provides good agreement with the fully quantum results, but it performs less well for weak pumping and in the middle of the transition region. Interestingly, the error of the method does not necessarily decay with the characteristic system scale as one could naively expect. Instead, higher photon numbers lead to sharper transitions which make the TWA less accurate. 

Finally, we studied the relaxation time as given by the Fokker-Planck evolution. According to our observations this leads to reasonable results in terms of the expected behavior, the extrapolation of previous values using numerical diagonalization \cite{Kahan_2021}, and the (very limited) comparison with results obtained from the quantum jumps formalism. We stress that while the TWA leads to results that are quantitatively not accurate, they are qualitatively correct, displaying a slowing down of the relaxation rates as the system approaches the semiclassical limit.

Our generalized semiclassical procedure represents a clear and inexpensive improvement over localized descriptions. It is a very fast form to obtain reliable estimators for the observables of interest, which can be useful as approximate predictions or as a starting point for more sophisticated and costly methods. We also note that the semiclassical approximation localized at the global minimum of an effective potential for the ion can provide a rather mistaken prediction for the location of the transition between different equilibrium configurations. This issue is especially relevant for systems with many local minima, such as variations of the Frenkel-Kontorova model \cite{fogarty2015, Buchheit_2020}. Even for problems for which the localized Gaussian description is considered appropriate, the method we use can provide a valuable benchmark.

The TWA is easily generalizable and still inexpensive to simulate systems with larger numbers of ions, where the long-ranged Coulomb interaction competes with the optical and trap forces producing richer dynamics \cite{fogarty2015, cormick2012structural}.  It could also be useful for the extension of previous work on the design of ion crystals \cite{laupretre2019} to the regime of higher cooperativities. Indeed, Eq.~\eqref{eq:TWA} can be modified to apply to a larger phase space including more ions. The use of the TWA gives, for the unitary evolution, a Poisson bracket in each pair of canonically conjugated variables. Losses or noise on the ions can be introduced by means of either independent or correlated drift and diffusion terms depending on the model of interest. Although phase-space representations with many degrees of freedom are very hard to visualize, the calculation of mean values is still fast, and the scaling with the number of ions is way less problematic than in a quantum treatment. As suggested by the results in this work the steady state obtained with a TWA could be significantly more accurate than the localized Gaussian approximation, with affordable computational resources.

\section*{Acknowledgments}

The authors acknowledge funding from Grants No. PICT 2017-2583, No. PICT 2018-02331, and No. PICT 2020-SERIEA-00959 from ANPCyT (Argentina). This work used computational resources from CCAD – Universidad Nacional de Córdoba, which are part of SNCAD – MinCyT, República Argentina.

\appendix

\section{Evolution of the fluctuations in the localized Gaussian approximation}
\label{sec:Gaussian}

In the following we provide the linearized system of equations governing the evolution of the fluctuations in the standard, localized Gaussian treatment. We refer to \cite{cormick2013} for an explanation of the details of the derivation. Continuing the procedure started in Sec. \ref{sec:standard semiclassical}, for each operator $A$ corresponding to cavity or motional quadratures we define fluctuation operators in the form $A=\overline A + \delta A$ with $\overline A$ the mean value associated with the semiclassical solution, and $\delta A$ the fluctuation operator with zero mean. 

We then derive Heisenberg-Langevin equations of motion for the fluctuations neglecting quadratic terms in any of the fluctuation operators:
\begin{equation}
    \begin{aligned}
        \dot {\delta a} & = \left( i \Delta_{\textrm{eff}}(\overline x)- \kappa \right)\delta a + i \frac{ c }{\sqrt{2}} (b + b^\dagger) + \sqrt{2 \kappa} a_{\textrm{in}}(t)\\
        \dot b & = \frac{i}{2\omega} \left( - \left( \omega^2 + \omega_v^2 \right) b + \left( \omega^2 - \omega_v^2 \right) b^\dagger \right) \\ 
        & \quad + i \frac{c}{\sqrt{2}} \left( e^{-i\varphi} \delta a + e^{i\varphi} \; \delta a^\dagger \right)
    \end{aligned}
    \label{eq:fluctuations}
\end{equation}

\noindent Here, $\delta a$ corresponds to fluctuations of the cavity operator $a$, and $b$, $b^\dagger$ are annihilation and creation operators respectively for the motional fluctuations of the ion, obtained combining $\delta x/x_\omega$ and $\delta p/p_\omega$. The phase factors $e^{\pm i\varphi}$ come from the phase of the mean cavity field:
\begin{equation}
\overline{a} = |\overline{a}| e^{i\varphi}
\end{equation}
and the coupling between motional and cavity fluctuations is given by:
\begin{equation}
    c = |\overline a| x_\omega \frac{d \Delta_{\textrm{eff}}(\overline  x)}{d \overline  x} \,.
\end{equation}
Finally, $a_{\textrm{in}}(t)$ represents the input operator within the input-output formalism \cite{gardiner_book,gardiner1985}, and
\begin{equation}
    \omega_v^2= \omega \left( \omega - |\overline a |^2  x_\omega^2 \frac{d^2 \Delta_{\textrm{eff}}(\overline x)}{d \overline x^2}\right)
\end{equation}
contains the correction to the vibrational frequency introduced by the optical potential to lowest order.

As only first-order terms in the fluctuations are kept, the evolution can be written in terms of Gaussian states at all times, and only the vector of first moments and the covariance matrix are needed to completely describe the system's state. 

We stress that the linearization implies that the effective detuning appearing in the set of equations \eqref{eq:fluctuations} is evaluated at the classical equilibrium position $\overline x$. Furthermore, the fluctuations of cavity and ion decouple when $\overline x$ corresponds to a minimum or maximum of the optical potential. Both issues can be a major source of error for the localized Gaussian approximation. The values of $\Delta_{\rm eff}$ and of the coupling constant $c$ are key aspects for the quality of cavity cooling \cite{cormick2013,Fogarty_2016}. The impact of the linearization is especially large when $\Delta_{\rm} (\overline x)$ is close to zero, within the bistable regime, and/or in absence of direct dissipation on the ion. We note that the particular error arising from using $\Delta_{\rm eff} (\overline x)$ in \eqref{eq:fluctuations} could be reduced by replacing it with a value of $\langle \Delta_{\rm eff}\rangle$ determined in a self-consistent manner.

\section{Numerical simulation of the Fokker-Planck evolution through stochastic differential equations}
\label{sec:stochastic}

The Fokker-Planck equation provides the dynamic description of a probability distribution through the solution of a second-order partial differential equation. To solve it we use a numerical approach: the solution of stochastic differential equations (SDE's) describing the underlying stochastic process. An n-dimensional vector satisfying the SDE

\begin{equation}
  dx = A (x,t) + B(x,t) dW_t \,,
\end{equation}

\noindent where $dW_t$ is a Wiener process, is equivalent to a Fokker-Planck equation \cite{gardiner1985handbook} when an ensemble of trajectories is evolved \cite{Huber_2021}. 

The equation then takes the form:
\begin{equation}\label{fokker_planck}
\begin{aligned}
  & \frac{\partial}{\partial t} p(\mathbf{x},t) = 
       - \sum_{i=1}^N \frac{\partial}{\partial x_i} \left( A_i(\mathbf{x},t) p(\mathbf{x},t) \right)\\
      & + \frac{1}{2}\sum_{i=1}^N \sum_{j=1}^N \frac{\partial^2}{\partial x_i \partial x_j} \left( B(\mathbf{x},t) B^T (\mathbf{x},t) p(\mathbf{x},t) \right)
\end{aligned}
\end{equation}
\noindent where $p$ is the probability density, $A$ is the drift and $D=B B^T$ is the diffusion matrix. The numerical solution of a system of stochastic differential equations can be accurately approximated by a variety of methods \cite{gardiner1985handbook, kloeden1994}. For our problem, an order $1.5$ Strong Taylor Scheme was chosen, following the implementation described in \cite{kloeden1994}.

\section{Semiclassical estimation of the relaxation rate based on overlaps}
\label{sec:overlaps}

As a first approach to estimate relaxation rates, we developed a simplified model that was conceived as a simplified version of previous studies of metastability \cite{rose2016, macieszczak2016, macieszczak2020}. Since obtaining a spectral decomposition of the Liouvillian is numerically prohibitive for the regime of interest, we defined a set of two semiclassical states which were used to approximate the metastable manifold. For simplicity, we took this set to be formed by combinations of Gaussian states. In particular, the equilibrium configuration corresponding to the ion located at the center was chosen as one Gaussian state, which we call $\rho_c$. The other relevant semiclassical configuration was chosen to be a statistical superposition $\rho_s = 1/2 (\rho_+ + \rho_-)$. Here, the subindices ``$c$'' and ``$s$'' refer to center and sides respectively, whereas $\rho_\pm$ are Gaussian states with the ion located at either side, with the subscript $\pm$ indicating the sign of the mean value of the position in each case. We note that the states involve also the degrees of freedom of the cavity, but we label them through the ion position for convenience. Each of the three Gaussian states $\rho_c, \rho_\pm$ was described by means of the corresponding first moments and covariance matrix \cite{Adesso2014}, which were found fitting the state obtained from FP evolution at times long enough to reach convergence. 

We then considered the action of the full evolution superoperator $\mathcal{L}$, without any truncations. It is straightforward to compute the action of this superoperator on the Gaussian states $\rho_c$ and $\rho_{\pm}$ expressed in phase space. In order to find the dynamics in the metastable manifold, one should project the results onto the left eigenvectors of $\mathcal{L}$ \cite{rose2016, macieszczak2016, macieszczak2020}. The left eigenvector corresponding to the null eigenvalue is the identity; however, the following left eigenvector is unknown. As an alternative approximate procedure, we formed a matrix projecting back onto the original semiclassical states, obtaining elements of the form $L_{jk}={\rm Tr}[\rho_j \mathcal{L} \rho_k]$, with $j,l \in \{c,s\}$. This again can be cast in terms of phase-space integrals. 

Naively, one could expect that the matrix containing these overlaps would allow one to approximate an effective evolution superoperator within the metastable manifold, in a similar spirit as \cite{rose2016, macieszczak2016, macieszczak2020}. This is unfortunately not the case, since the Gaussian approximations we find are not close enough to true metastable states. In the first place, the FP evolution from which we extract the metastable manifold is truncated, and secondly, the states obtained from FP evolution are generally not close to Gaussians. This results in very large diagonal terms $L_{jj}$. These terms, which should be very small if the states are approximately stable, are actually larger than the tunneling (off-diagonal) terms by many orders of magnitude. Even worse, the diagonal terms turned out to be positive, so they could not be associated with a loss of population due to tunneling out of the given configuration.

Finding an accurate description of a metastable non-Gaussian semiclassical state can become as hard as solving the full problem, so we resorted to a different strategy: we ignored the diagonal terms in this matrix and considered only the off-diagonal terms, to check whether $|L_{jk}|$ with $j\neq k$ could provide a reasonable estimate of the tunneling rates between semiclassical configurations. However, this led to exceedingly small relaxation rates, several orders of magnitude smaller than the ones expected from extrapolating from the regime where numerical diagonalization is feasible. This suggests a failure of this method, possibly because the non-Gaussian features of the localized states are crucial for the description of the tunneling between configurations.

\section{Delimitation of individual states in phase space}
\label{sec:delimitation}

\begin{figure}[h]
\centering
      \includegraphics[width=1\columnwidth]{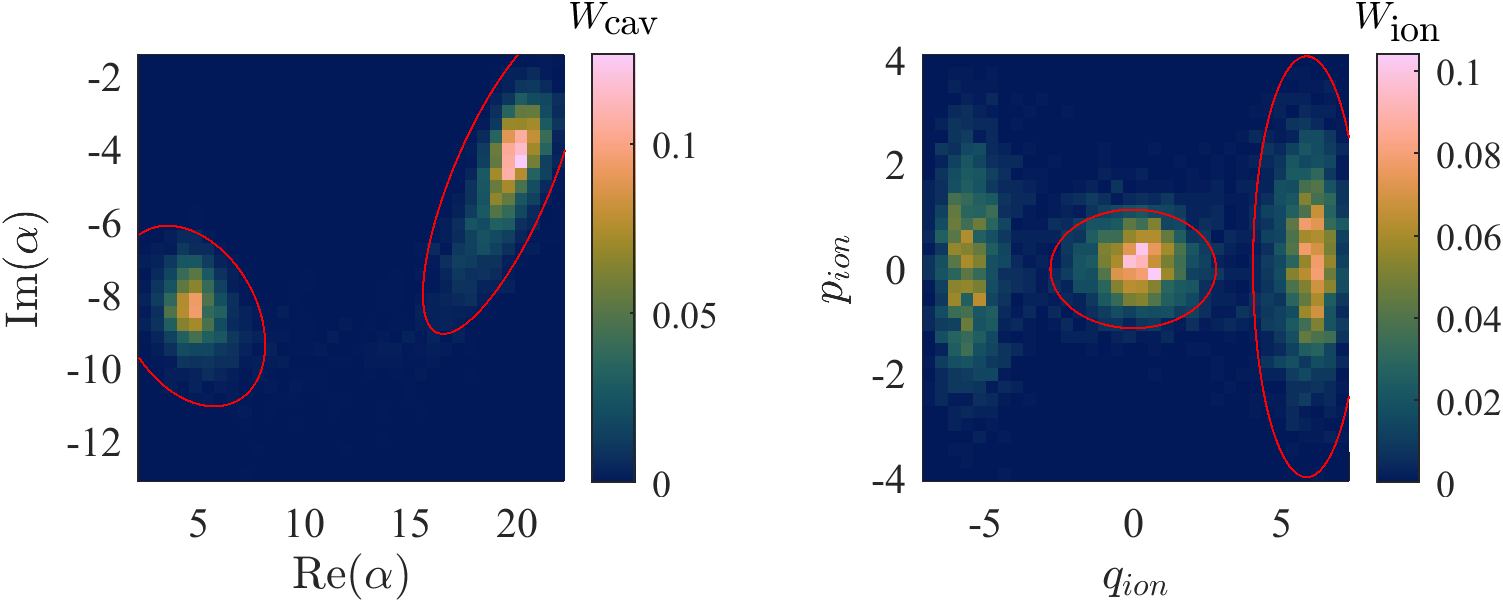}
  \caption{Wigner distribution obtained through TWA. The left and right columns show the Wigner distribution of the intracavity field and the ion, respectively. The red lines indicate the regions defined for the two-state stochastic model.}
  \label{fig:comp_regiones_cav_ion}
\end{figure}

In the semiclassically bistable region, the phase space representation of the steady state is bimodal (Fig. \ref{fig:comp_regiones_cav_ion}), with each peak corresponding to different classical equilibrium configurations. We delimited boundaries for these distributions so we could distinguish between them and estimate the corresponding transition rates.
To define their boundaries, a Gaussian mixture model was fitted to the steady state distribution functions of each subsystem.
This steady state was found as the long-time evolution of a normally distributed initial condition, centered approximately at the unstable equilibrium position of the effective potential [Eq. (\ref{eq:Veff})].
Then, each of the Gaussian components was used without their relative weights as initial probability distributions for the subsequent evolution of the FP equation, thus allowing the estimation of the mean first passage time towards each state. 

The regions that delimited each state in phase space were chosen as ellipses centered at each of the two-dimensional vectors of mean values. The semiaxes of such ellipses were defined as 3 standard deviations in the principal directions of each covariance matrix. For the shaded areas in Fig.~\ref{fig:FP_tunneling_rates}, we redefined the ellipses using 2.5 and 3.5 standard deviations (instead of 3) to check how much this impacted on the results. The shaded areas enclose the region between the minimum and the maximum rates found in the three cases.

We remark that only the first escape time was used in the calculation of the first passage time since after reentering the initial region, the probability density would not be normally distributed and the meaning of subsequent escape times is not clear. When the initial distribution was fixed at the sides, the former was sampled from both the left and right (ion position) Gaussian distributions.


%


\begin{thebibliography}{46}%
\makeatletter
\providecommand \@ifxundefined [1]{%
 \@ifx{#1\undefined}
}%
\providecommand \@ifnum [1]{%
 \ifnum #1\expandafter \@firstoftwo
 \else \expandafter \@secondoftwo
 \fi
}%
\providecommand \@ifx [1]{%
 \ifx #1\expandafter \@firstoftwo
 \else \expandafter \@secondoftwo
 \fi
}%
\providecommand \natexlab [1]{#1}%
\providecommand \enquote  [1]{``#1''}%
\providecommand \bibnamefont  [1]{#1}%
\providecommand \bibfnamefont [1]{#1}%
\providecommand \citenamefont [1]{#1}%
\providecommand \href@noop [0]{\@secondoftwo}%
\providecommand \href [0]{\begingroup \@sanitize@url \@href}%
\providecommand \@href[1]{\@@startlink{#1}\@@href}%
\providecommand \@@href[1]{\endgroup#1\@@endlink}%
\providecommand \@sanitize@url [0]{\catcode `\\12\catcode `\$12\catcode
  `\&12\catcode `\#12\catcode `\^12\catcode `\_12\catcode `\%12\relax}%
\providecommand \@@startlink[1]{}%
\providecommand \@@endlink[0]{}%
\providecommand \url  [0]{\begingroup\@sanitize@url \@url }%
\providecommand \@url [1]{\endgroup\@href {#1}{\urlprefix }}%
\providecommand \urlprefix  [0]{URL }%
\providecommand \Eprint [0]{\href }%
\providecommand \doibase [0]{http://dx.doi.org/}%
\providecommand \selectlanguage [0]{\@gobble}%
\providecommand \bibinfo  [0]{\@secondoftwo}%
\providecommand \bibfield  [0]{\@secondoftwo}%
\providecommand \translation [1]{[#1]}%
\providecommand \BibitemOpen [0]{}%
\providecommand \bibitemStop [0]{}%
\providecommand \bibitemNoStop [0]{.\EOS\space}%
\providecommand \EOS [0]{\spacefactor3000\relax}%
\providecommand \BibitemShut  [1]{\csname bibitem#1\endcsname}%
\let\auto@bib@innerbib\@empty
\bibitem [{\citenamefont {Altman}\ \emph {et~al.}(2021)\citenamefont {Altman},
  \citenamefont {Brown}, \citenamefont {Carleo}, \citenamefont {Carr},
  \citenamefont {Demler}, \citenamefont {Chin}, \citenamefont {DeMarco},
  \citenamefont {Economou}, \citenamefont {Eriksson}, \citenamefont {Fu},
  \citenamefont {Greiner}, \citenamefont {Hazzard}, \citenamefont {Hulet},
  \citenamefont {Koll\'ar}, \citenamefont {Lev}, \citenamefont {Lukin},
  \citenamefont {Ma}, \citenamefont {Mi}, \citenamefont {Misra}, \citenamefont
  {Monroe}, \citenamefont {Murch}, \citenamefont {Nazario}, \citenamefont {Ni},
  \citenamefont {Potter}, \citenamefont {Roushan}, \citenamefont {Saffman},
  \citenamefont {Schleier-Smith}, \citenamefont {Siddiqi}, \citenamefont
  {Simmonds}, \citenamefont {Singh}, \citenamefont {Spielman}, \citenamefont
  {Temme}, \citenamefont {Weiss}, \citenamefont {Vu\ifmmode \check{c}\else
  \v{c}\fi{}kovi\ifmmode~\acute{c}\else \'{c}\fi{}}, \citenamefont
  {Vuleti\ifmmode~\acute{c}\else \'{c}\fi{}}, \citenamefont {Ye},\ and\
  \citenamefont {Zwierlein}}]{Altman_2021}%
  \BibitemOpen
  \bibfield  {author} {\bibinfo {author} {\bibfnamefont {E.}~\bibnamefont
  {Altman}}, \bibinfo {author} {\bibfnamefont {K.~R.}\ \bibnamefont {Brown}},
  \bibinfo {author} {\bibfnamefont {G.}~\bibnamefont {Carleo}}, \bibinfo
  {author} {\bibfnamefont {L.~D.}\ \bibnamefont {Carr}}, \bibinfo {author}
  {\bibfnamefont {E.}~\bibnamefont {Demler}}, \bibinfo {author} {\bibfnamefont
  {C.}~\bibnamefont {Chin}}, \bibinfo {author} {\bibfnamefont {B.}~\bibnamefont
  {DeMarco}}, \bibinfo {author} {\bibfnamefont {S.~E.}\ \bibnamefont
  {Economou}}, \bibinfo {author} {\bibfnamefont {M.~A.}\ \bibnamefont
  {Eriksson}}, \bibinfo {author} {\bibfnamefont {K.-M.~C.}\ \bibnamefont {Fu}}
  {\it et al.},\ }\href {\doibase
  10.1103/PRXQuantum.2.017003} {\bibfield  {journal} {\bibinfo  {journal} {PRX
  Quantum}\ }\textbf {\bibinfo {volume} {2}},\ \bibinfo {pages} {017003}
  (\bibinfo {year} {2021})}\BibitemShut {NoStop}%
\bibitem [{\citenamefont {Porras}\ and\ \citenamefont
  {Cirac}(2004)}]{Porras_2004}%
  \BibitemOpen
  \bibfield  {author} {\bibinfo {author} {\bibfnamefont {D.}~\bibnamefont
  {Porras}}\ and\ \bibinfo {author} {\bibfnamefont {J.~I.}\ \bibnamefont
  {Cirac}},\ }\href {\doibase 10.1103/PhysRevLett.92.207901} {\bibfield
  {journal} {\bibinfo  {journal} {Phys. Rev. Lett.}\ }\textbf {\bibinfo
  {volume} {92}},\ \bibinfo {pages} {207901} (\bibinfo {year}
  {2004})}\BibitemShut {NoStop}%
\bibitem [{\citenamefont {Monroe}\ \emph {et~al.}(2021)\citenamefont {Monroe},
  \citenamefont {Campbell}, \citenamefont {Duan}, \citenamefont {Gong},
  \citenamefont {Gorshkov}, \citenamefont {Hess}, \citenamefont {Islam},
  \citenamefont {Kim}, \citenamefont {Linke}, \citenamefont {Pagano},
  \citenamefont {Richerme}, \citenamefont {Senko},\ and\ \citenamefont
  {Yao}}]{Monroe_2021}%
  \BibitemOpen
  \bibfield  {author} {\bibinfo {author} {\bibfnamefont {C.}~\bibnamefont
  {Monroe}}, \bibinfo {author} {\bibfnamefont {W.~C.}\ \bibnamefont
  {Campbell}}, \bibinfo {author} {\bibfnamefont {L.-M.}\ \bibnamefont {Duan}},
  \bibinfo {author} {\bibfnamefont {Z.-X.}\ \bibnamefont {Gong}}, \bibinfo
  {author} {\bibfnamefont {A.~V.}\ \bibnamefont {Gorshkov}}, \bibinfo {author}
  {\bibfnamefont {P.~W.}\ \bibnamefont {Hess}}, \bibinfo {author}
  {\bibfnamefont {R.}~\bibnamefont {Islam}}, \bibinfo {author} {\bibfnamefont
  {K.}~\bibnamefont {Kim}}, \bibinfo {author} {\bibfnamefont {N.~M.}\
  \bibnamefont {Linke}}, \bibinfo {author} {\bibfnamefont {G.}~\bibnamefont
  {Pagano}}, \bibinfo {author} {\bibfnamefont {P.}~\bibnamefont {Richerme}},
  \bibinfo {author} {\bibfnamefont {C.}~\bibnamefont {Senko}}, \ and\ \bibinfo
  {author} {\bibfnamefont {N.~Y.}\ \bibnamefont {Yao}},\ }\href {\doibase
  10.1103/RevModPhys.93.025001} {\bibfield  {journal} {\bibinfo  {journal}
  {Rev. Mod. Phys.}\ }\textbf {\bibinfo {volume} {93}},\ \bibinfo {pages}
  {025001} (\bibinfo {year} {2021})}\BibitemShut {NoStop}%
\bibitem [{\citenamefont {Bermudez}\ \emph {et~al.}(2013)\citenamefont
  {Bermudez}, \citenamefont {Bruderer},\ and\ \citenamefont
  {Plenio}}]{Bermudez_2013}%
  \BibitemOpen
  \bibfield  {author} {\bibinfo {author} {\bibfnamefont {A.}~\bibnamefont
  {Bermudez}}, \bibinfo {author} {\bibfnamefont {M.}~\bibnamefont {Bruderer}},
  \ and\ \bibinfo {author} {\bibfnamefont {M.~B.}\ \bibnamefont {Plenio}},\
  }\href {\doibase 10.1103/PhysRevLett.111.040601} {\bibfield  {journal}
  {\bibinfo  {journal} {Phys. Rev. Lett.}\ }\textbf {\bibinfo {volume} {111}},\
  \bibinfo {pages} {040601} (\bibinfo {year} {2013})}\BibitemShut {NoStop}%
\bibitem [{\citenamefont {Cormick}\ and\ \citenamefont
  {Schmiegelow}(2016)}]{Cormick_PRA_2016}%
  \BibitemOpen
  \bibfield  {author} {\bibinfo {author} {\bibfnamefont {C.}~\bibnamefont
  {Cormick}}\ and\ \bibinfo {author} {\bibfnamefont {C.~T.}\ \bibnamefont
  {Schmiegelow}},\ }\href {\doibase 10.1103/PhysRevA.94.053406} {\bibfield
  {journal} {\bibinfo  {journal} {Phys. Rev. A}\ }\textbf {\bibinfo {volume}
  {94}},\ \bibinfo {pages} {053406} (\bibinfo {year} {2016})}\BibitemShut
  {NoStop}%
\bibitem [{\citenamefont {Mezzacapo}\ \emph {et~al.}(2012)\citenamefont
  {Mezzacapo}, \citenamefont {Casanova}, \citenamefont {Lamata},\ and\
  \citenamefont {Solano}}]{Mezzacapo_2012}%
  \BibitemOpen
  \bibfield  {author} {\bibinfo {author} {\bibfnamefont {A.}~\bibnamefont
  {Mezzacapo}}, \bibinfo {author} {\bibfnamefont {J.}~\bibnamefont {Casanova}},
  \bibinfo {author} {\bibfnamefont {L.}~\bibnamefont {Lamata}}, \ and\ \bibinfo
  {author} {\bibfnamefont {E.}~\bibnamefont {Solano}},\ }\href {\doibase
  10.1103/PhysRevLett.109.200501} {\bibfield  {journal} {\bibinfo  {journal}
  {Phys. Rev. Lett.}\ }\textbf {\bibinfo {volume} {109}},\ \bibinfo {pages}
  {200501} (\bibinfo {year} {2012})}\BibitemShut {NoStop}%
\bibitem [{\citenamefont {Lemmer}\ \emph {et~al.}(2018)\citenamefont {Lemmer},
  \citenamefont {Cormick}, \citenamefont {Tamascelli}, \citenamefont {Schaetz},
  \citenamefont {Huelga},\ and\ \citenamefont {Plenio}}]{Lemmer_2018}%
  \BibitemOpen
  \bibfield  {author} {\bibinfo {author} {\bibfnamefont {A.}~\bibnamefont
  {Lemmer}}, \bibinfo {author} {\bibfnamefont {C.}~\bibnamefont {Cormick}},
  \bibinfo {author} {\bibfnamefont {D.}~\bibnamefont {Tamascelli}}, \bibinfo
  {author} {\bibfnamefont {T.}~\bibnamefont {Schaetz}}, \bibinfo {author}
  {\bibfnamefont {S.~F.}\ \bibnamefont {Huelga}}, \ and\ \bibinfo {author}
  {\bibfnamefont {M.~B.}\ \bibnamefont {Plenio}},\ }\href {\doibase
  10.1088/1367-2630/aac87d} {\bibfield  {journal} {\bibinfo  {journal} {New J.
  Phys.}\ }\textbf {\bibinfo {volume} {20}},\ \bibinfo {pages} {073002}
  (\bibinfo {year} {2018})}\BibitemShut {NoStop}%
\bibitem [{\citenamefont {Pachos}\ and\ \citenamefont
  {Walther}(2002)}]{Pachos_2002}%
  \BibitemOpen
  \bibfield  {author} {\bibinfo {author} {\bibfnamefont {J.}~\bibnamefont
  {Pachos}}\ and\ \bibinfo {author} {\bibfnamefont {H.}~\bibnamefont
  {Walther}},\ }\href {\doibase 10.1103/PhysRevLett.89.187903} {\bibfield
  {journal} {\bibinfo  {journal} {Phys. Rev. Lett.}\ }\textbf {\bibinfo
  {volume} {89}},\ \bibinfo {pages} {187903} (\bibinfo {year}
  {2002})}\BibitemShut {NoStop}%
\bibitem [{\citenamefont {Schmiegelow}\ \emph {et~al.}(2016)\citenamefont
  {Schmiegelow}, \citenamefont {Kaufmann}, \citenamefont {Ruster},
  \citenamefont {Schulz}, \citenamefont {Kaushal}, \citenamefont {Hettrich},
  \citenamefont {Schmidt-Kaler},\ and\ \citenamefont
  {Poschinger}}]{Schmiegelow_2016}%
  \BibitemOpen
  \bibfield  {author} {\bibinfo {author} {\bibfnamefont {C.~T.}\ \bibnamefont
  {Schmiegelow}}, \bibinfo {author} {\bibfnamefont {H.}~\bibnamefont
  {Kaufmann}}, \bibinfo {author} {\bibfnamefont {T.}~\bibnamefont {Ruster}},
  \bibinfo {author} {\bibfnamefont {J.}~\bibnamefont {Schulz}}, \bibinfo
  {author} {\bibfnamefont {V.}~\bibnamefont {Kaushal}}, \bibinfo {author}
  {\bibfnamefont {M.}~\bibnamefont {Hettrich}}, \bibinfo {author}
  {\bibfnamefont {F.}~\bibnamefont {Schmidt-Kaler}}, \ and\ \bibinfo {author}
  {\bibfnamefont {U.~G.}\ \bibnamefont {Poschinger}},\ }\href {\doibase
  10.1103/PhysRevLett.116.033002} {\bibfield  {journal} {\bibinfo  {journal}
  {Phys. Rev. Lett.}\ }\textbf {\bibinfo {volume} {116}},\ \bibinfo {pages}
  {033002} (\bibinfo {year} {2016})}\BibitemShut {NoStop}%
\bibitem [{\citenamefont {Linnet}\ \emph
  {et~al.}(2012{\natexlab{a}})\citenamefont {Linnet}, \citenamefont {Leroux},
  \citenamefont {Marciante}, \citenamefont {Dantan},\ and\ \citenamefont
  {Drewsen}}]{Linnet_2012}%
  \BibitemOpen
  \bibfield  {author} {\bibinfo {author} {\bibfnamefont {R.~B.}\ \bibnamefont
  {Linnet}}, \bibinfo {author} {\bibfnamefont {I.~D.}\ \bibnamefont {Leroux}},
  \bibinfo {author} {\bibfnamefont {M.}~\bibnamefont {Marciante}}, \bibinfo
  {author} {\bibfnamefont {A.}~\bibnamefont {Dantan}}, \ and\ \bibinfo {author}
  {\bibfnamefont {M.}~\bibnamefont {Drewsen}},\ }\href {\doibase
  10.1103/PhysRevLett.109.233005} {\bibfield  {journal} {\bibinfo  {journal}
  {Phys. Rev. Lett.}\ }\textbf {\bibinfo {volume} {109}},\ \bibinfo {pages}
  {233005} (\bibinfo {year} {2012}{\natexlab{a}})}\BibitemShut {NoStop}%
\bibitem [{\citenamefont {Pruttivarasin}\ \emph {et~al.}(2011)\citenamefont
  {Pruttivarasin}, \citenamefont {Ramm}, \citenamefont {Talukdar},
  \citenamefont {Kreuter},\ and\ \citenamefont
  {H\"affner}}]{Pruttivarasin2011}%
  \BibitemOpen
  \bibfield  {author} {\bibinfo {author} {\bibfnamefont {T.}~\bibnamefont
  {Pruttivarasin}}, \bibinfo {author} {\bibfnamefont {M.}~\bibnamefont {Ramm}},
  \bibinfo {author} {\bibfnamefont {I.}~\bibnamefont {Talukdar}}, \bibinfo
  {author} {\bibfnamefont {A.}~\bibnamefont {Kreuter}}, \ and\ \bibinfo
  {author} {\bibfnamefont {H.}~\bibnamefont {H\"affner}},\ }\href
  {http://stacks.iop.org/1367-2630/13/i=7/a=075012} {\bibfield  {journal}
  {\bibinfo  {journal} {New J. Phys.}\ }\textbf {\bibinfo {volume} {13}},\
  \bibinfo {pages} {075012} (\bibinfo {year} {2011})}\BibitemShut {NoStop}%
\bibitem [{\citenamefont {Ramette}\ \emph {et~al.}(2022)\citenamefont
  {Ramette}, \citenamefont {Sinclair}, \citenamefont {Vendeiro}, \citenamefont
  {Rudelis}, \citenamefont {Cetina},\ and\ \citenamefont
  {Vuleti\ifmmode~\acute{c}\else \'{c}\fi{}}}]{Ramette_2022}%
  \BibitemOpen
  \bibfield  {author} {\bibinfo {author} {\bibfnamefont {J.}~\bibnamefont
  {Ramette}}, \bibinfo {author} {\bibfnamefont {J.}~\bibnamefont {Sinclair}},
  \bibinfo {author} {\bibfnamefont {Z.}~\bibnamefont {Vendeiro}}, \bibinfo
  {author} {\bibfnamefont {A.}~\bibnamefont {Rudelis}}, \bibinfo {author}
  {\bibfnamefont {M.}~\bibnamefont {Cetina}}, \ and\ \bibinfo {author}
  {\bibfnamefont {V.}~\bibnamefont {Vuleti\ifmmode~\acute{c}\else
  \'{c}\fi{}}},\ }\href {\doibase 10.1103/PRXQuantum.3.010344} {\bibfield
  {journal} {\bibinfo  {journal} {PRX Quantum}\ }\textbf {\bibinfo {volume}
  {3}},\ \bibinfo {pages} {010344} (\bibinfo {year} {2022})}\BibitemShut
  {NoStop}%
\bibitem [{\citenamefont {Schmied}\ \emph {et~al.}(2008)\citenamefont
  {Schmied}, \citenamefont {Roscilde}, \citenamefont {Murg}, \citenamefont
  {Porras},\ and\ \citenamefont {Cirac}}]{Schmied_2008}%
  \BibitemOpen
  \bibfield  {author} {\bibinfo {author} {\bibfnamefont {R.}~\bibnamefont
  {Schmied}}, \bibinfo {author} {\bibfnamefont {T.}~\bibnamefont {Roscilde}},
  \bibinfo {author} {\bibfnamefont {V.}~\bibnamefont {Murg}}, \bibinfo {author}
  {\bibfnamefont {D.}~\bibnamefont {Porras}}, \ and\ \bibinfo {author}
  {\bibfnamefont {J.~I.}\ \bibnamefont {Cirac}},\ }\href {\doibase
  10.1088/1367-2630/10/4/045017} {\bibfield  {journal} {\bibinfo  {journal}
  {New J. Phys.}\ }\textbf {\bibinfo {volume} {10}},\ \bibinfo {pages} {045017}
  (\bibinfo {year} {2008})}\BibitemShut {NoStop}%
\bibitem [{\citenamefont {Fogarty}\ \emph {et~al.}(2016)\citenamefont
  {Fogarty}, \citenamefont {Landa}, \citenamefont {Cormick},\ and\
  \citenamefont {Morigi}}]{Fogarty_2016}%
  \BibitemOpen
  \bibfield  {author} {\bibinfo {author} {\bibfnamefont {T.}~\bibnamefont
  {Fogarty}}, \bibinfo {author} {\bibfnamefont {H.}~\bibnamefont {Landa}},
  \bibinfo {author} {\bibfnamefont {C.}~\bibnamefont {Cormick}}, \ and\
  \bibinfo {author} {\bibfnamefont {G.}~\bibnamefont {Morigi}},\ }\href
  {\doibase 10.1103/PhysRevA.94.023844} {\bibfield  {journal} {\bibinfo
  {journal} {Phys. Rev. A}\ }\textbf {\bibinfo {volume} {94}},\ \bibinfo
  {pages} {023844} (\bibinfo {year} {2016})}\BibitemShut {NoStop}%
\bibitem [{\citenamefont {Cormick}\ and\ \citenamefont
  {Morigi}(2012)}]{cormick2012structural}%
  \BibitemOpen
  \bibfield  {author} {\bibinfo {author} {\bibfnamefont {C.}~\bibnamefont
  {Cormick}}\ and\ \bibinfo {author} {\bibfnamefont {G.}~\bibnamefont
  {Morigi}},\ }\href {\doibase 10.1103/PhysRevLett.109.053003} {\bibfield
  {journal} {\bibinfo  {journal} {Phys. Rev. Lett.}\ }\textbf {\bibinfo
  {volume} {109}},\ \bibinfo {pages} {053003} (\bibinfo {year}
  {2012})}\BibitemShut {NoStop}%
\bibitem [{\citenamefont {Cormick}\ \emph {et~al.}(2011)\citenamefont
  {Cormick}, \citenamefont {Schaetz},\ and\ \citenamefont
  {Morigi}}]{Cormick_2011}%
  \BibitemOpen
  \bibfield  {author} {\bibinfo {author} {\bibfnamefont {C.}~\bibnamefont
  {Cormick}}, \bibinfo {author} {\bibfnamefont {T.}~\bibnamefont {Schaetz}}, \
  and\ \bibinfo {author} {\bibfnamefont {G.}~\bibnamefont {Morigi}},\ }\href
  {\doibase 10.1088/1367-2630/13/4/043019} {\bibfield  {journal} {\bibinfo
  {journal} {New J. Phys.}\ }\textbf {\bibinfo {volume} {13}},\ \bibinfo
  {pages} {043019} (\bibinfo {year} {2011})}\BibitemShut {NoStop}%
\bibitem [{\citenamefont {Garcia-Mata}\ \emph {et~al.}(2007)\citenamefont
  {Garcia-Mata}, \citenamefont {Zhirov},\ and\ \citenamefont
  {Shepelyansky}}]{Garcia-Mata2007}%
  \BibitemOpen
  \bibfield  {author} {\bibinfo {author} {\bibfnamefont {I.}~\bibnamefont
  {Garcia-Mata}}, \bibinfo {author} {\bibfnamefont {O.~V.}\ \bibnamefont
  {Zhirov}}, \ and\ \bibinfo {author} {\bibfnamefont {D.~L.}\ \bibnamefont
  {Shepelyansky}},\ }\href {\doibase 10.1140/epjd/e2006-00220-2} {\bibfield
  {journal} {\bibinfo  {journal} {Eur. Phys. J. D}\ }\textbf {\bibinfo {volume}
  {41}},\ \bibinfo {pages} {325} (\bibinfo {year} {2007})}\BibitemShut
  {NoStop}%
\bibitem [{\citenamefont {Bylinskii}\ \emph {et~al.}(2015)\citenamefont
  {Bylinskii}, \citenamefont {Gangloff},\ and\ \citenamefont
  {Vuleti{\'{c}}}}]{bylinskii2015}%
  \BibitemOpen
  \bibfield  {author} {\bibinfo {author} {\bibfnamefont {A.}~\bibnamefont
  {Bylinskii}}, \bibinfo {author} {\bibfnamefont {D.}~\bibnamefont {Gangloff}},
  \ and\ \bibinfo {author} {\bibfnamefont {V.}~\bibnamefont {Vuleti{\'{c}}}},\
  }\href {\doibase 10.1126/science.1261422} {\bibfield  {journal} {\bibinfo
  {journal} {Science}\ }\textbf {\bibinfo {volume} {348}},\ \bibinfo {pages}
  {1115} (\bibinfo {year} {2015})}\BibitemShut {NoStop}%
\bibitem [{\citenamefont {Bylinskii}\ \emph {et~al.}(2016)\citenamefont
  {Bylinskii}, \citenamefont {Gangloff}, \citenamefont {Counts},\ and\
  \citenamefont {Vuleti{\'c}}}]{bylinskii2016observation}%
  \BibitemOpen
  \bibfield  {author} {\bibinfo {author} {\bibfnamefont {A.}~\bibnamefont
  {Bylinskii}}, \bibinfo {author} {\bibfnamefont {D.}~\bibnamefont {Gangloff}},
  \bibinfo {author} {\bibfnamefont {I.}~\bibnamefont {Counts}}, \ and\ \bibinfo
  {author} {\bibfnamefont {V.}~\bibnamefont {Vuleti{\'c}}},\ }\href {\doibase
  10.1038/nmat4601} {\bibfield  {journal} {\bibinfo  {journal} {Nature Mat.}\
  }\textbf {\bibinfo {volume} {15}},\ \bibinfo {pages} {717} (\bibinfo {year}
  {2016})}\BibitemShut {NoStop}%
\bibitem [{\citenamefont {Fogarty}\ \emph {et~al.}(2015)\citenamefont
  {Fogarty}, \citenamefont {Cormick}, \citenamefont {Landa}, \citenamefont
  {Stojanovi{\'{c}}}, \citenamefont {Demler},\ and\ \citenamefont
  {Morigi}}]{fogarty2015}%
  \BibitemOpen
  \bibfield  {author} {\bibinfo {author} {\bibfnamefont {T.}~\bibnamefont
  {Fogarty}}, \bibinfo {author} {\bibfnamefont {C.}~\bibnamefont {Cormick}},
  \bibinfo {author} {\bibfnamefont {H.}~\bibnamefont {Landa}}, \bibinfo
  {author} {\bibfnamefont {V.~M.}\ \bibnamefont {Stojanovi{\'{c}}}}, \bibinfo
  {author} {\bibfnamefont {E.}~\bibnamefont {Demler}}, \ and\ \bibinfo {author}
  {\bibfnamefont {G.}~\bibnamefont {Morigi}},\ }\href {\doibase
  10.1103/PhysRevLett.115.233602} {\bibfield  {journal} {\bibinfo  {journal}
  {Phys. Rev. Lett.}\ }\textbf {\bibinfo {volume} {115}},\ \bibinfo {pages}
  {233602} (\bibinfo {year} {2015})}\BibitemShut {NoStop}%
\bibitem [{\citenamefont {Kahan}\ \emph {et~al.}(2021)\citenamefont {Kahan},
  \citenamefont {Ermann},\ and\ \citenamefont {Cormick}}]{Kahan_2021}%
  \BibitemOpen
  \bibfield  {author} {\bibinfo {author} {\bibfnamefont {A.}~\bibnamefont
  {Kahan}}, \bibinfo {author} {\bibfnamefont {L.}~\bibnamefont {Ermann}}, \
  and\ \bibinfo {author} {\bibfnamefont {C.}~\bibnamefont {Cormick}},\ }\href
  {\doibase 10.1103/PhysRevA.104.043705} {\bibfield  {journal} {\bibinfo
  {journal} {Phys. Rev. A}\ }\textbf {\bibinfo {volume} {104}},\ \bibinfo
  {pages} {043705} (\bibinfo {year} {2021})}\BibitemShut {NoStop}%
\bibitem [{\citenamefont {Carmichael}(1993)}]{Carmichael_1993_book}%
  \BibitemOpen
  \bibfield  {author} {\bibinfo {author} {\bibfnamefont {H.}~\bibnamefont
  {Carmichael}},\ }\href
  {https://link.springer.com/book/10.1007/978-3-540-47620-7} { {\bibinfo
  {title} {An Open Systems Approach to Quantum Optics}}}\ (\bibinfo
  {publisher} {Springer Berlin Heidelberg},\ \bibinfo {year}
  {1993})\BibitemShut {NoStop}%
\bibitem [{\citenamefont {Gardiner}\ and\ \citenamefont
  {Zoller}(2000)}]{gardiner_book}%
  \BibitemOpen
  \bibfield  {author} {\bibinfo {author} {\bibfnamefont {C.~W.}\ \bibnamefont
  {Gardiner}}\ and\ \bibinfo {author} {\bibfnamefont {P.}~\bibnamefont
  {Zoller}},\ }\href {https://link.springer.com/book/9783540223016} {
  {\bibinfo {title} {Quantum Noise}}},\ edited by\ \bibinfo {editor}
  {\bibfnamefont {H.}~\bibnamefont {Haken}}\ (\bibinfo  {publisher}
  {Springer},\ \bibinfo {year} {2000})\BibitemShut {NoStop}%
\bibitem [{\citenamefont {Verstraelen}\ \emph {et~al.}(2020)\citenamefont
  {Verstraelen}, \citenamefont {Rota}, \citenamefont {Savona},\ and\
  \citenamefont {Wouters}}]{Verstraelen_2020}%
  \BibitemOpen
  \bibfield  {author} {\bibinfo {author} {\bibfnamefont {W.}~\bibnamefont
  {Verstraelen}}, \bibinfo {author} {\bibfnamefont {R.}~\bibnamefont {Rota}},
  \bibinfo {author} {\bibfnamefont {V.}~\bibnamefont {Savona}}, \ and\ \bibinfo
  {author} {\bibfnamefont {M.}~\bibnamefont {Wouters}},\ }\href {\doibase
  10.1103/PhysRevResearch.2.022037} {\bibfield  {journal} {\bibinfo  {journal}
  {Phys. Rev. Res.}\ }\textbf {\bibinfo {volume} {2}},\ \bibinfo {pages}
  {022037 (R)} (\bibinfo {year} {2020})}\BibitemShut {NoStop}%
\bibitem [{\citenamefont {Huber}\ \emph {et~al.}(2021)\citenamefont {Huber},
  \citenamefont {Kirton},\ and\ \citenamefont {Rabl}}]{Huber_2021}%
  \BibitemOpen
  \bibfield  {author} {\bibinfo {author} {\bibfnamefont {J.}~\bibnamefont
  {Huber}}, \bibinfo {author} {\bibfnamefont {P.}~\bibnamefont {Kirton}}, \
  and\ \bibinfo {author} {\bibfnamefont {P.}~\bibnamefont {Rabl}},\ }\href
  {\doibase 10.21468/SciPostPhys.10.2.045} {\bibfield  {journal} {\bibinfo
  {journal} {SciPost Phys.}\ }\textbf {\bibinfo {volume} {10}},\ \bibinfo
  {pages} {45} (\bibinfo {year} {2021})}\BibitemShut {NoStop}%
\bibitem [{\citenamefont {Lescanne}\ \emph {et~al.}(2019)\citenamefont
  {Lescanne}, \citenamefont {Verney}, \citenamefont {Ficheux}, \citenamefont
  {Devoret}, \citenamefont {Huard}, \citenamefont {Mirrahimi},\ and\
  \citenamefont {Leghtas}}]{lescanne2019}%
  \BibitemOpen
  \bibfield  {author} {\bibinfo {author} {\bibfnamefont {R.}~\bibnamefont
  {Lescanne}}, \bibinfo {author} {\bibfnamefont {L.}~\bibnamefont {Verney}},
  \bibinfo {author} {\bibfnamefont {Q.}~\bibnamefont {Ficheux}}, \bibinfo
  {author} {\bibfnamefont {M.~H.}\ \bibnamefont {Devoret}}, \bibinfo {author}
  {\bibfnamefont {B.}~\bibnamefont {Huard}}, \bibinfo {author} {\bibfnamefont
  {M.}~\bibnamefont {Mirrahimi}}, \ and\ \bibinfo {author} {\bibfnamefont
  {Z.}~\bibnamefont {Leghtas}},\ }\href {\doibase
  10.1103/PhysRevApplied.11.014030} {\bibfield  {journal} {\bibinfo  {journal}
  {Phys. Rev. Appl.}\ }\textbf {\bibinfo {volume} {11}},\ \bibinfo {pages}
  {014030} (\bibinfo {year} {2019})}\BibitemShut {NoStop}%
\bibitem [{\citenamefont {Vicentini}\ \emph {et~al.}(2018)\citenamefont
  {Vicentini}, \citenamefont {Minganti}, \citenamefont {Rota}, \citenamefont
  {Orso},\ and\ \citenamefont {Ciuti}}]{vicentini2018}%
  \BibitemOpen
  \bibfield  {author} {\bibinfo {author} {\bibfnamefont {F.}~\bibnamefont
  {Vicentini}}, \bibinfo {author} {\bibfnamefont {F.}~\bibnamefont {Minganti}},
  \bibinfo {author} {\bibfnamefont {R.}~\bibnamefont {Rota}}, \bibinfo {author}
  {\bibfnamefont {G.}~\bibnamefont {Orso}}, \ and\ \bibinfo {author}
  {\bibfnamefont {C.}~\bibnamefont {Ciuti}},\ }\href {\doibase
  10.1103/PhysRevA.97.013853} {\bibfield  {journal} {\bibinfo  {journal} {Phys.
  Rev. A}\ }\textbf {\bibinfo {volume} {97}},\ \bibinfo {pages} {013853}
  (\bibinfo {year} {2018})}\BibitemShut {NoStop}%
\bibitem [{\citenamefont {Hwang}\ \emph {et~al.}(2018)\citenamefont {Hwang},
  \citenamefont {Rabl},\ and\ \citenamefont {Plenio}}]{hwang2018}%
  \BibitemOpen
  \bibfield  {author} {\bibinfo {author} {\bibfnamefont {M.-J.}\ \bibnamefont
  {Hwang}}, \bibinfo {author} {\bibfnamefont {P.}~\bibnamefont {Rabl}}, \ and\
  \bibinfo {author} {\bibfnamefont {M.~B.}\ \bibnamefont {Plenio}},\ }\href
  {\doibase 10.1103/PhysRevA.97.013825} {\bibfield  {journal} {\bibinfo
  {journal} {Phys. Rev. A}\ }\textbf {\bibinfo {volume} {97}},\ \bibinfo
  {pages} {013825} (\bibinfo {year} {2018})}\BibitemShut {NoStop}%
\bibitem [{\citenamefont {Dettmer}\ \emph {et~al.}(2001)\citenamefont
  {Dettmer}, \citenamefont {Hellweg}, \citenamefont {Ryytty}, \citenamefont
  {Arlt}, \citenamefont {Ertmer}, \citenamefont {Sengstock}, \citenamefont
  {Petrov}, \citenamefont {Shlyapnikov}, \citenamefont {Kreutzmann},
  \citenamefont {Santos},\ and\ \citenamefont {Lewenstein}}]{dettmer2001}%
  \BibitemOpen
  \bibfield  {author} {\bibinfo {author} {\bibfnamefont {S.}~\bibnamefont
  {Dettmer}}, \bibinfo {author} {\bibfnamefont {D.}~\bibnamefont {Hellweg}},
  \bibinfo {author} {\bibfnamefont {P.}~\bibnamefont {Ryytty}}, \bibinfo
  {author} {\bibfnamefont {J.~J.}\ \bibnamefont {Arlt}}, \bibinfo {author}
  {\bibfnamefont {W.}~\bibnamefont {Ertmer}}, \bibinfo {author} {\bibfnamefont
  {K.}~\bibnamefont {Sengstock}}, \bibinfo {author} {\bibfnamefont {D.~S.}\
  \bibnamefont {Petrov}}, \bibinfo {author} {\bibfnamefont {G.~V.}\
  \bibnamefont {Shlyapnikov}}, \bibinfo {author} {\bibfnamefont
  {H.}~\bibnamefont {Kreutzmann}}, \bibinfo {author} {\bibfnamefont
  {L.}~\bibnamefont {Santos}}, \ and\ \bibinfo {author} {\bibfnamefont
  {M.}~\bibnamefont {Lewenstein}},\ }\href {\doibase
  10.1103/PhysRevLett.87.160406} {\bibfield  {journal} {\bibinfo  {journal}
  {Phys. Rev. Lett.}\ }\textbf {\bibinfo {volume} {87}},\ \bibinfo {pages}
  {160406} (\bibinfo {year} {2001})}\BibitemShut {NoStop}%
\bibitem [{\citenamefont {Zhang}\ and\ \citenamefont
  {Baranger}(2021)}]{zhang2021}%
  \BibitemOpen
  \bibfield  {author} {\bibinfo {author} {\bibfnamefont {X.~H.~H.}\
  \bibnamefont {Zhang}}\ and\ \bibinfo {author} {\bibfnamefont {H.~U.}\
  \bibnamefont {Baranger}},\ }\href {\doibase 10.1103/PhysRevA.103.033711}
  {\bibfield  {journal} {\bibinfo  {journal} {Phys. Rev. A}\ }\textbf {\bibinfo
  {volume} {103}},\ \bibinfo {pages} {033711} (\bibinfo {year}
  {2021})}\BibitemShut {NoStop}%
\bibitem [{\citenamefont {Ritsch}\ \emph {et~al.}(2013)\citenamefont {Ritsch},
  \citenamefont {Domokos}, \citenamefont {Brennecke},\ and\ \citenamefont
  {Esslinger}}]{ritsch2013}%
  \BibitemOpen
  \bibfield  {author} {\bibinfo {author} {\bibfnamefont {H.}~\bibnamefont
  {Ritsch}}, \bibinfo {author} {\bibfnamefont {P.}~\bibnamefont {Domokos}},
  \bibinfo {author} {\bibfnamefont {F.}~\bibnamefont {Brennecke}}, \ and\
  \bibinfo {author} {\bibfnamefont {T.}~\bibnamefont {Esslinger}},\ }\href
  {\doibase 10.1103/RevModPhys.85.553} {\bibfield  {journal} {\bibinfo
  {journal} {Rev. Mod. Phys.}\ }\textbf {\bibinfo {volume} {85}},\ \bibinfo
  {pages} {553} (\bibinfo {year} {2013})}\BibitemShut {NoStop}%
\bibitem [{\citenamefont {Cormick}\ and\ \citenamefont
  {Morigi}(2013)}]{cormick2013}%
  \BibitemOpen
  \bibfield  {author} {\bibinfo {author} {\bibfnamefont {C.}~\bibnamefont
  {Cormick}}\ and\ \bibinfo {author} {\bibfnamefont {G.}~\bibnamefont
  {Morigi}},\ }\href {\doibase 10.1103/PhysRevA.87.013829} {\bibfield
  {journal} {\bibinfo  {journal} {Phys. Rev. A}\ }\textbf {\bibinfo {volume}
  {87}},\ \bibinfo {pages} {013829} (\bibinfo {year} {2013})}\BibitemShut
  {NoStop}%
\bibitem [{\citenamefont {Maunz}\ \emph {et~al.}(2001)\citenamefont {Maunz},
  \citenamefont {Fischer}, \citenamefont {Puppe}, \citenamefont {Pinkse},\ and\
  \citenamefont {Rempe}}]{maunz2001}%
  \BibitemOpen
  \bibfield  {author} {\bibinfo {author} {\bibfnamefont {P.}~\bibnamefont
  {Maunz}}, \bibinfo {author} {\bibfnamefont {T.}~\bibnamefont {Fischer}},
  \bibinfo {author} {\bibfnamefont {T.}~\bibnamefont {Puppe}}, \bibinfo
  {author} {\bibfnamefont {P.~W.}\ \bibnamefont {Pinkse}}, \ and\ \bibinfo
  {author} {\bibfnamefont {G.}~\bibnamefont {Rempe}},\ }\href {\doibase
  10.1109/QELS.2001.961902} {\bibfield  {journal} {\bibinfo  {journal}
  {Technical Digest - Summaries of Papers Presented at the Quantum Electronics
  and Laser Science Conference, QELS 2001}\ ,\ \bibinfo {pages} {94}} (\bibinfo
  {year} {2001})}\BibitemShut {NoStop}%
\bibitem [{\citenamefont {Verstraelen}\ and\ \citenamefont
  {Wouters}(2018)}]{verstraelen2018}%
  \BibitemOpen
  \bibfield  {author} {\bibinfo {author} {\bibfnamefont {W.}~\bibnamefont
  {Verstraelen}}\ and\ \bibinfo {author} {\bibfnamefont {M.}~\bibnamefont
  {Wouters}},\ }\href {\doibase 10.3390/app8091427} {\bibfield  {journal}
  {\bibinfo  {journal} {Appl. Sci.}\ }\textbf {\bibinfo {volume} {8}},\
  \bibinfo {pages} {1427} (\bibinfo {year} {2018})}\BibitemShut {NoStop}%
\bibitem [{\citenamefont {Schleich}(2001)}]{Schleich_book}%
  \BibitemOpen
  \bibfield  {author} {\bibinfo {author} {\bibfnamefont {W.~P.}\ \bibnamefont
  {Schleich}},\ }\href {\doibase 10.1002/3527602976} { {\bibinfo {title}
  {Quantum Optics in Phase Space}}}\ (\bibinfo  {publisher} {John Wiley \&
  Sons, Ltd},\ \bibinfo {year} {2001})\BibitemShut {NoStop}%
\bibitem [{\citenamefont {Brack}\ and\ \citenamefont
  {Bhaduri}(2018)}]{brack_2018}%
  \BibitemOpen
  \bibfield  {author} {\bibinfo {author} {\bibfnamefont {M.}~\bibnamefont
  {Brack}}\ and\ \bibinfo {author} {\bibfnamefont {R.~K.}\ \bibnamefont
  {Bhaduri}},\ }\href {\doibase 10.1201/9780429502828} { {\bibinfo {title}
  {Semiclassical physics}}}\ (\bibinfo  {publisher} {CRC Press},\ \bibinfo
  {year} {2018})\BibitemShut {NoStop}%
\bibitem [{\citenamefont {Reichl}(1992)}]{reichl_1992}%
  \BibitemOpen
  \bibfield  {author} {\bibinfo {author} {\bibfnamefont {L.~E.}\ \bibnamefont
  {Reichl}},\ }\href@noop {} { {\bibinfo {title} {The transition to
  chaos}}},\ Vol.~\bibinfo {volume} {6}\ (\bibinfo  {publisher} {Springer},\
  \bibinfo {year} {1992})\BibitemShut {NoStop}%
\bibitem [{\citenamefont {Schneider}\ \emph {et~al.}(2010)\citenamefont
  {Schneider}, \citenamefont {Enderlein}, \citenamefont {Huber},\ and\
  \citenamefont {Schaetz}}]{schneider2010}%
  \BibitemOpen
  \bibfield  {author} {\bibinfo {author} {\bibfnamefont {C.}~\bibnamefont
  {Schneider}}, \bibinfo {author} {\bibfnamefont {M.}~\bibnamefont
  {Enderlein}}, \bibinfo {author} {\bibfnamefont {T.}~\bibnamefont {Huber}}, \
  and\ \bibinfo {author} {\bibfnamefont {T.}~\bibnamefont {Schaetz}},\ }\href
  {\doibase 10.1038/nphoton.2010.236} {\bibfield  {journal} {\bibinfo
  {journal} {Nature Phot.}\ }\textbf {\bibinfo {volume} {4}},\ \bibinfo {pages}
  {772} (\bibinfo {year} {2010})}\BibitemShut {NoStop}%
\bibitem [{\citenamefont {Steiner}\ \emph {et~al.}(2013)\citenamefont
  {Steiner}, \citenamefont {Meyer}, \citenamefont {Deutsch}, \citenamefont
  {Reichel},\ and\ \citenamefont {K\"ohl}}]{steiner2013}%
  \BibitemOpen
  \bibfield  {author} {\bibinfo {author} {\bibfnamefont {M.}~\bibnamefont
  {Steiner}}, \bibinfo {author} {\bibfnamefont {H.~M.}\ \bibnamefont {Meyer}},
  \bibinfo {author} {\bibfnamefont {C.}~\bibnamefont {Deutsch}}, \bibinfo
  {author} {\bibfnamefont {J.}~\bibnamefont {Reichel}}, \ and\ \bibinfo
  {author} {\bibfnamefont {M.}~\bibnamefont {K\"ohl}},\ }\href {\doibase
  10.1103/PhysRevLett.110.043003} {\bibfield  {journal} {\bibinfo  {journal}
  {Phys. Rev. Lett.}\ }\textbf {\bibinfo {volume} {110}},\ \bibinfo {pages}
  {043003} (\bibinfo {year} {2013})}\BibitemShut {NoStop}%
\bibitem [{\citenamefont {Thompson}\ \emph {et~al.}(1992)\citenamefont
  {Thompson}, \citenamefont {Rempe},\ and\ \citenamefont
  {Kimble}}]{thompson1992}%
  \BibitemOpen
  \bibfield  {author} {\bibinfo {author} {\bibfnamefont {R.~J.}\ \bibnamefont
  {Thompson}}, \bibinfo {author} {\bibfnamefont {G.}~\bibnamefont {Rempe}}, \
  and\ \bibinfo {author} {\bibfnamefont {H.~J.}\ \bibnamefont {Kimble}},\
  }\href {\doibase 10.1103/PhysRevLett.68.1132} {\bibfield  {journal} {\bibinfo
   {journal} {Phys. Rev. Lett.}\ }\textbf {\bibinfo {volume} {68}},\ \bibinfo
  {pages} {1132} (\bibinfo {year} {1992})}\BibitemShut {NoStop}%
\bibitem [{\citenamefont {Herskind}\ \emph {et~al.}(2009)\citenamefont
  {Herskind}, \citenamefont {Dantan}, \citenamefont {Marler}, \citenamefont
  {Albert},\ and\ \citenamefont {Drewsen}}]{herskind2009}%
  \BibitemOpen
  \bibfield  {author} {\bibinfo {author} {\bibfnamefont {P.~F.}\ \bibnamefont
  {Herskind}}, \bibinfo {author} {\bibfnamefont {A.}~\bibnamefont {Dantan}},
  \bibinfo {author} {\bibfnamefont {J.~P.}\ \bibnamefont {Marler}}, \bibinfo
  {author} {\bibfnamefont {M.}~\bibnamefont {Albert}}, \ and\ \bibinfo {author}
  {\bibfnamefont {M.}~\bibnamefont {Drewsen}},\ }\href {\doibase
  10.1038/nphys1302} {\bibfield  {journal} {\bibinfo  {journal} {Nature Phys.}\
  }\textbf {\bibinfo {volume} {5}},\ \bibinfo {pages} {494} (\bibinfo {year}
  {2009})}\BibitemShut {NoStop}%
\bibitem [{\citenamefont {Lee}\ \emph {et~al.}(2019)\citenamefont {Lee},
  \citenamefont {Friebe}, \citenamefont {Fioretto}, \citenamefont
  {Sch\"uppert}, \citenamefont {Ong}, \citenamefont {Plankensteiner},
  \citenamefont {Torggler}, \citenamefont {Ritsch}, \citenamefont {Blatt},\
  and\ \citenamefont {Northup}}]{lee2018}%
  \BibitemOpen
  \bibfield  {author} {\bibinfo {author} {\bibfnamefont {M.}~\bibnamefont
  {Lee}}, \bibinfo {author} {\bibfnamefont {K.}~\bibnamefont {Friebe}},
  \bibinfo {author} {\bibfnamefont {D.~A.}\ \bibnamefont {Fioretto}}, \bibinfo
  {author} {\bibfnamefont {K.}~\bibnamefont {Sch\"uppert}}, \bibinfo {author}
  {\bibfnamefont {F.~R.}\ \bibnamefont {Ong}}, \bibinfo {author} {\bibfnamefont
  {D.}~\bibnamefont {Plankensteiner}}, \bibinfo {author} {\bibfnamefont
  {V.}~\bibnamefont {Torggler}}, \bibinfo {author} {\bibfnamefont
  {H.}~\bibnamefont {Ritsch}}, \bibinfo {author} {\bibfnamefont
  {R.}~\bibnamefont {Blatt}}, \ and\ \bibinfo {author} {\bibfnamefont {T.~E.}\
  \bibnamefont {Northup}},\ }\href {\doibase 10.1103/PhysRevLett.122.153603}
  {\bibfield  {journal} {\bibinfo  {journal} {Phys. Rev. Lett.}\ }\textbf
  {\bibinfo {volume} {122}},\ \bibinfo {pages} {153603} (\bibinfo {year}
  {2019})}\BibitemShut {NoStop}%
\bibitem [{\citenamefont {Meraner}\ \emph {et~al.}(2020)\citenamefont
  {Meraner}, \citenamefont {Mazloom}, \citenamefont {Krutyanskiy},
  \citenamefont {Krcmarsky}, \citenamefont {Schupp}, \citenamefont {Fioretto},
  \citenamefont {Sekatski}, \citenamefont {Northup}, \citenamefont
  {Sangouard},\ and\ \citenamefont {Lanyon}}]{meraner2020}%
  \BibitemOpen
  \bibfield  {author} {\bibinfo {author} {\bibfnamefont {M.}~\bibnamefont
  {Meraner}}, \bibinfo {author} {\bibfnamefont {A.}~\bibnamefont {Mazloom}},
  \bibinfo {author} {\bibfnamefont {V.}~\bibnamefont {Krutyanskiy}}, \bibinfo
  {author} {\bibfnamefont {V.}~\bibnamefont {Krcmarsky}}, \bibinfo {author}
  {\bibfnamefont {J.}~\bibnamefont {Schupp}}, \bibinfo {author} {\bibfnamefont
  {D.~A.}\ \bibnamefont {Fioretto}}, \bibinfo {author} {\bibfnamefont
  {P.}~\bibnamefont {Sekatski}}, \bibinfo {author} {\bibfnamefont {T.~E.}\
  \bibnamefont {Northup}}, \bibinfo {author} {\bibfnamefont {N.}~\bibnamefont
  {Sangouard}}, \ and\ \bibinfo {author} {\bibfnamefont {B.~P.}\ \bibnamefont
  {Lanyon}},\ }\href {\doibase 10.1103/PhysRevA.102.052614} {\bibfield
  {journal} {\bibinfo  {journal} {Phys. Rev. A}\ }\textbf {\bibinfo {volume}
  {102}},\ \bibinfo {pages} {052614} (\bibinfo {year} {2020})}\BibitemShut
  {NoStop}%
\bibitem [{\citenamefont {Buchheit}\ and\ \citenamefont
  {Rjasanow}(2020)}]{Buchheit_2020}%
  \BibitemOpen
  \bibfield  {author} {\bibinfo {author} {\bibfnamefont {A.~A.}\ \bibnamefont
  {Buchheit}}\ and\ \bibinfo {author} {\bibfnamefont {S.}~\bibnamefont
  {Rjasanow}},\ }\href {\doibase https://doi.org/10.1016/j.physd.2019.132298}
  {\bibfield  {journal} {\bibinfo  {journal} {Physica D}\ }\textbf {\bibinfo
  {volume} {406}},\ \bibinfo {pages} {132298} (\bibinfo {year}
  {2020})}\BibitemShut {NoStop}%
\bibitem [{\citenamefont {Rose}\ \emph {et~al.}(2016)\citenamefont {Rose},
  \citenamefont {Macieszczak}, \citenamefont {Lesanovsky},\ and\ \citenamefont
  {Garrahan}}]{rose2016}%
  \BibitemOpen
  \bibfield  {author} {\bibinfo {author} {\bibfnamefont {D.~C.}\ \bibnamefont
  {Rose}}, \bibinfo {author} {\bibfnamefont {K.}~\bibnamefont {Macieszczak}},
  \bibinfo {author} {\bibfnamefont {I.}~\bibnamefont {Lesanovsky}}, \ and\
  \bibinfo {author} {\bibfnamefont {J.~P.}\ \bibnamefont {Garrahan}},\ }\href
  {\doibase 10.1103/PhysRevE.94.052132} {\bibfield  {journal} {\bibinfo
  {journal} {Phys. Rev. E}\ }\textbf {\bibinfo {volume} {94}},\ \bibinfo
  {pages} {052132} (\bibinfo {year} {2016})}\BibitemShut {NoStop}%
\bibitem [{\citenamefont {Macieszczak}\ \emph {et~al.}(2016)\citenamefont
  {Macieszczak}, \citenamefont {Gu{\c{t}}{\u{a}}}, \citenamefont {Lesanovsky},\
  and\ \citenamefont {Garrahan}}]{macieszczak2016}%
  \BibitemOpen
  \bibfield  {author} {\bibinfo {author} {\bibfnamefont {K.}~\bibnamefont
  {Macieszczak}}, \bibinfo {author} {\bibfnamefont {M.}~\bibnamefont
  {Gu{\c{t}}{\u{a}}}}, \bibinfo {author} {\bibfnamefont {I.}~\bibnamefont
  {Lesanovsky}}, \ and\ \bibinfo {author} {\bibfnamefont {J.~P.}\ \bibnamefont
  {Garrahan}},\ }\href {\doibase 10.1103/PhysRevLett.116.240404} {\bibfield
  {journal} {\bibinfo  {journal} {Phys. Rev. Lett.}\ }\textbf {\bibinfo
  {volume} {116}},\ \bibinfo {pages} {240404} (\bibinfo {year}
  {2016})}\BibitemShut {NoStop}%
\bibitem [{\citenamefont {Macieszczak}\ \emph {et~al.}(2021)\citenamefont
  {Macieszczak}, \citenamefont {Rose}, \citenamefont {Lesanovsky},\ and\
  \citenamefont {Garrahan}}]{macieszczak2020}%
  \BibitemOpen
  \bibfield  {author} {\bibinfo {author} {\bibfnamefont {K.}~\bibnamefont
  {Macieszczak}}, \bibinfo {author} {\bibfnamefont {D.~C.}\ \bibnamefont
  {Rose}}, \bibinfo {author} {\bibfnamefont {I.}~\bibnamefont {Lesanovsky}}, \
  and\ \bibinfo {author} {\bibfnamefont {J.~P.}\ \bibnamefont {Garrahan}},\
  }\href {\doibase 10.1103/physrevresearch.3.033047} {\bibfield  {journal}
  {\bibinfo  {journal} {Phys. Rev. Res.}\ }\textbf {\bibinfo {volume} {3}},\
  \bibinfo {pages} {033047} (\bibinfo {year} {2021})}\BibitemShut {NoStop}%
\bibitem [{\citenamefont {Okushima}\ \emph {et~al.}(2019)\citenamefont
  {Okushima}, \citenamefont {Niiyama}, \citenamefont {Ikeda},\ and\
  \citenamefont {Shimizu}}]{okushima2019}%
  \BibitemOpen
  \bibfield  {author} {\bibinfo {author} {\bibfnamefont {T.}~\bibnamefont
  {Okushima}}, \bibinfo {author} {\bibfnamefont {T.}~\bibnamefont {Niiyama}},
  \bibinfo {author} {\bibfnamefont {K.~S.}\ \bibnamefont {Ikeda}}, \ and\
  \bibinfo {author} {\bibfnamefont {Y.}~\bibnamefont {Shimizu}},\ }\href
  {\doibase 10.1103/PhysRevE.100.032311} {\bibfield  {journal} {\bibinfo
  {journal} {Phys. Rev. E}\ }\textbf {\bibinfo {volume} {100}},\ \bibinfo
  {pages} {032311} (\bibinfo {year} {2019})}\BibitemShut {NoStop}%
\bibitem [{\citenamefont {Kells}\ \emph {et~al.}(2019)\citenamefont {Kells},
  \citenamefont {Mih\'alka}, \citenamefont {Annibale},\ and\ \citenamefont
  {Rosta}}]{Kells_2019}%
  \BibitemOpen
  \bibfield  {author} {\bibinfo {author} {\bibfnamefont {A.}~\bibnamefont
  {Kells}}, \bibinfo {author} {\bibfnamefont {Z.~E.}\ \bibnamefont
  {Mih\'alka}}, \bibinfo {author} {\bibfnamefont {A.}~\bibnamefont {Annibale}},
  \ and\ \bibinfo {author} {\bibfnamefont {E.}~\bibnamefont {Rosta}},\ }\href
  {\doibase 10.1063/1.5083924} {\bibfield  {journal} {\bibinfo  {journal} {J.
  Chem. Phys.}\ }\textbf {\bibinfo {volume} {150}},\ \bibinfo {pages} {134107}
  (\bibinfo {year} {2019})}\BibitemShut {NoStop}%
\bibitem [{\citenamefont {Dalibard}\ \emph {et~al.}(1992)\citenamefont
  {Dalibard}, \citenamefont {Castin},\ and\ \citenamefont
  {M\o{}lmer}}]{Dalibard_1992}%
  \BibitemOpen
  \bibfield  {author} {\bibinfo {author} {\bibfnamefont {J.}~\bibnamefont
  {Dalibard}}, \bibinfo {author} {\bibfnamefont {Y.}~\bibnamefont {Castin}}, \
  and\ \bibinfo {author} {\bibfnamefont {K.}~\bibnamefont {M\o{}lmer}},\ }\href
  {\doibase 10.1103/PhysRevLett.68.580} {\bibfield  {journal} {\bibinfo
  {journal} {Phys. Rev. Lett.}\ }\textbf {\bibinfo {volume} {68}},\ \bibinfo
  {pages} {580} (\bibinfo {year} {1992})}\BibitemShut {NoStop}%
\bibitem [{\citenamefont {Laupr\^etre}\ \emph {et~al.}(2019)\citenamefont
  {Laupr\^etre}, \citenamefont {Linnet}, \citenamefont {Leroux}, \citenamefont
  {Landa}, \citenamefont {Dantan},\ and\ \citenamefont
  {Drewsen}}]{laupretre2019}%
  \BibitemOpen
  \bibfield  {author} {\bibinfo {author} {\bibfnamefont {T.}~\bibnamefont
  {Laupr\^etre}}, \bibinfo {author} {\bibfnamefont {R.~B.}\ \bibnamefont
  {Linnet}}, \bibinfo {author} {\bibfnamefont {I.~D.}\ \bibnamefont {Leroux}},
  \bibinfo {author} {\bibfnamefont {H.}~\bibnamefont {Landa}}, \bibinfo
  {author} {\bibfnamefont {A.}~\bibnamefont {Dantan}}, \ and\ \bibinfo {author}
  {\bibfnamefont {M.}~\bibnamefont {Drewsen}},\ }\href {\doibase
  10.1103/PhysRevA.99.031401} {\bibfield  {journal} {\bibinfo  {journal} {Phys.
  Rev. A}\ }\textbf {\bibinfo {volume} {99}},\ \bibinfo {pages} {031401}
  (\bibinfo {year} {2019})}\BibitemShut {NoStop}%
\bibitem [{\citenamefont {Gardiner}\ and\ \citenamefont
  {Collett}(1985)}]{gardiner1985}%
  \BibitemOpen
  \bibfield  {author} {\bibinfo {author} {\bibfnamefont {C.~W.}\ \bibnamefont
  {Gardiner}}\ and\ \bibinfo {author} {\bibfnamefont {M.~J.}\ \bibnamefont
  {Collett}},\ }\href {\doibase 10.1103/PhysRevA.31.3761} {\bibfield  {journal}
  {\bibinfo  {journal} {Phys. Rev. A}\ }\textbf {\bibinfo {volume} {31}},\
  \bibinfo {pages} {3761} (\bibinfo {year} {1985})}\BibitemShut {NoStop}%
\bibitem [{\citenamefont {Gardiner}\ \emph {et~al.}(1985)\citenamefont
  {Gardiner}, \citenamefont {Schurz}}]{gardiner1985handbook}%
  \BibitemOpen
  \bibfield  {author} {\bibinfo {author} {\bibfnamefont {C.~W.}~\bibnamefont
  {Gardiner}},\
  }\href {https://books.google.com.ar/books?id=cRfvAAAAMAAJ} {
  {\bibinfo  {title} {Handbook of Stochastic Methods for Physics, Chemistry, and the Natural Sciences}\ }\ }\ (\bibinfo {publisher} {Springer-Verlag},\ \bibinfo {year} {1985})\BibitemShut {NoStop}%
\bibitem [{\citenamefont {Kloeden}\ \emph {et~al.}(1994)\citenamefont
  {Platen}, \citenamefont {Schurz}}]{kloeden1994}%
  \BibitemOpen
  \bibfield  {author} {\bibinfo {author} {\bibfnamefont {P.~E.}~\bibnamefont
  {Kloeden}}, \bibinfo {author} {\bibfnamefont {E.}~\bibnamefont {Platen}},
  \ and\ \bibinfo {author} {\bibfnamefont {H.}\ \bibnamefont {Schurz}},\
  }\href {\doibase 10.1007/978-3-642-57913-4_4} {
  {\bibinfo  {title} {Numerical Solution of SDE Through Computer Experiments}\ }\ }\ (\bibinfo {publisher} {Springer Berlin Heidelberg},\ \bibinfo {year} {1994})\BibitemShut {NoStop}%
\bibitem [{\citenamefont {{Adesso}}\ \emph {et~al.}(2014)\citenamefont
  {{Adesso}}, \citenamefont {{Ragy}},\ and\ \citenamefont
  {{Lee}}}]{Adesso2014}%
  \BibitemOpen
  \bibfield  {author} {\bibinfo {author} {\bibfnamefont {G.}~\bibnamefont
  {{Adesso}}}, \bibinfo {author} {\bibfnamefont {S.}~\bibnamefont {{Ragy}}}, \
  and\ \bibinfo {author} {\bibfnamefont {A.~R.}\ \bibnamefont {{Lee}}},\ }\href
  {\doibase 10.1142/S1230161214400010} {\bibfield  {journal} {\bibinfo
  {journal} {Open Syst. Inf. Dyn.}\ }\textbf {\bibinfo {volume} {21}},\
  \bibinfo {pages} {1440001} (\bibinfo {year} {2014})}\BibitemShut {NoStop}%
\end{thebibliography}
\end{document}